\documentclass{aa}
\usepackage{graphicx}
\usepackage{txfonts}
\usepackage[]{natbib}
\usepackage[]{longtable}
\usepackage{multirow}

\bibpunct{(}{)}{;}{a}{}{,}

\begin{document}
   \title{Spectral catalogue of bright gamma--ray bursts detected with the BeppoSAX/GRBM}


   \author{C.~Guidorzi\inst{1}, M.~Lacapra\inst{1}, F. Frontera\inst{1,2}, E.~Montanari\inst{1,3},
   L.~Amati\inst{2}, F.~Calura\inst{4}, L.~Nicastro\inst{2}, M.~Orlandini\inst{2}
}

   \offprints{C.\ Guidorzi, guidorzi@fe.infn.it}
   \institute{Dipartimento di Fisica, Universit\`a di Ferrara, via Saragat 1, I-44122 Ferrara, Italy
   \and INAF--IASF Bologna, via P. Gobetti 101, I-40129 Bologna, Italy
   \and Istituto IS Calvi, Finale Emilia (MO), Italy
   \and Jeremiah Horrocks Institute for Astrophysics and Supercomputing, University of Central Lancashire, Preston PR1 2HE, UK}

  \date{Received 14 September 2010 / Accepted 23 October 2010}
%
\abstract
{The emission process responsible for the so--called ``prompt'' emission of gamma--ray bursts is still
unknown. A number of empirical models fitting the typical spectrum still lack a satisfactory interpretation.
A few GRB spectral catalogues derived from past and present experiments are known in the literature and
allow to tackle the issue of spectral properties of gamma--ray bursts on a statistical ground.}
{We extracted and studied the time--integrated photon spectra of the 200 brightest GRBs observed with the
Gamma--Ray Burst Monitor which flew aboard the BeppoSAX mission (1996--2002) to provide an independent
statistical characterisation of GRB spectra.}
{The spectra have a time-resolution of 128~s and consist of 240~energy channels covering the 40--700~keV energy
band. The 200 brightest GRBs were selected from the complete catalogue of 1082 GRBs detected
with the GRBM (Frontera et~al. 2009), whose products are publicly available and can be browsed/retrieved using a
dedicated web interface.
The spectra were fit with three models: a simple power--law,  a cut--off power law or a Band model.
We derived the sample distributions of the best-fitting spectral parameters and investigated possible
correlations between them. For a few, typically very long GRBs, we also provide a loose (128-s) time--resolved
spectroscopic analysis.}
{The typical photon spectrum of a bright GRB consists of a low--energy index around $1.0$  and a peak
energy of the $\nu\,F_\nu$ spectrum $E_{\rm p}\simeq 240$~keV in agreement with previous results on a sample of
bright CGRO/BATSE bursts.
Spectra of $\sim 35$\% of GRBs can be fit with a power--law with a photon index around 2, indicative
of peak energies either close to or outside the GRBM energy boundaries.
We confirm the correlation between $E_{\rm p}$ and fluence, in agreement with previous results,
with a logarithmic dispersion of $0.13$ around the power--law with index $0.21 \pm 0.06$. This is
shallower than its analogous in the GRB rest--frame, the Amati relation, between the intrinsic peak energy
and the isotropic--equivalent released energy (slope of $\sim 0.5$). The reason for this difference
mainly lies in the instrumental selection effect connected with the finite energy range of the GRBM
particularly at low energies.}
{We confirm the statistical properties of the low--energy and peak energy distributions found by other
experiments. These properties are not yet systematically explained in the current literature with the
proposed emission processes. The capability of measuring time--resolved spectra over a broadband energy range,
ensuring precise measurements of parameters such as $E_{\rm p}$, will be of key importance for future experiments.}

{}
\authorrunning {C. Guidorzi  et al.}
\titlerunning {GRB spectral catalogue of BeppoSAX/GRBM}
\maketitle

\section{Introduction}
\label{sec:intro}
Giant leaps in the knowledge of the gamma-ray burst (GRB) explosions have
been made in the last 15 years, mainly thanks to the discoveries obtained by
former BeppoSAX (1996--2002), HETE--II (2000--2006) and current Swift (2004) and
Fermi (2008) missions, as well as those made by ground facilities in response to
the spacecraft triggers.

Time-integrated photon spectra of long GRBs can be
adequately fit with a smoothly broken power--law \citep{Band93}, whose
low-energy and high-energy photon indices, $\alpha$ and $\beta$,
have median values of $-1$ and $-2.3$, respectively
(\citealt{Preece00,Kaneko06}; hereafter K06).
Similar results were obtained by time-resolved spectral analysis 
(\citealt{Frontera00,Ghirlanda02}; K06).
In spite of this, the nature and emission mechanisms responsible for the
prompt emission of GRBs are still a matter of debate.

The corresponding $\nu$$F_\nu$ spectrum peaks at $E_{\rm p}$,
the so-called peak energy, whose rest-frame value is found to correlate
with other relevant observed intrinsic properties, such as the
isotropic-equivalent radiated $\gamma$-ray energy, $E_{\rm iso}$
\citep{Amati02},
or its collimation-corrected value, $E_{\gamma}$ \citep{Ghirlanda04}. 
These correlations are observed to hold statistically on the sample of GRBs
with known intrinsic quantities; however, they are affected by a significant
dispersion, which could be due to some hidden variables. Specifically, while
the scatter of the $E_{\rm p}$--$E_{\rm iso}$ relation is well measured and known
to differ from zero (e.g., \citealt{Amati09}), the same issue for the corresponding
collimation--corrected relation is debated (e.g., \citealt{Campana07,Ghirlanda07,McBreen10}).
In the BATSE catalogue \citep{Paciesas99}, the $E_{\rm p}$ distribution clusters
around $300$~keV with a $\sim 100$~keV width (K06).

From the phenomenological perspective much 
effort has been made in order to characterise and identify typical spectral properties 
of bursts, by applying parametric spectral models 
that characterise, within the observational energy
window, the most relevant quantities. These quantities include the peak energy 
and the low and high energy components which are related,
according to the most accredited emission theories, 
to the particle energy distribution and/or to the physical parameters of
the emitting region. 

One of the most promising mechanisms proposed for the gamma--ray emission is the
synchrotron shock model (SSM). This model assumes that the electrons in an optically
thin environment are accelerated by the first--order Fermi mechanism to a power--law
distribution $dN(\gamma)/d\gamma \propto \gamma^{-p}$, where $\gamma$ is the Lorentz factor.
This distribution does not evolve in time, and the electron index $p$ is related to
the high--energy photon index either as $\beta = - (p + 2) / 2$ in case of
the fast--cooling synchrotron spectrum, or as $\beta = - (p + 1) / 2$  in case of
non-cooling synchrotron \citep{Sari98}.
The peak energy can be expressed as: $E_{\rm p} \propto \gamma_{e}^{2} B_{ps} $, where
$ \gamma_{e}^{2} $ is the pre--shock  equilibrium electron energy and $B_{ps}$ is the
post--shock magnetic field  \citep{Tavani96}. 

The use of large catalogues represents a fruitful approach to the study of the
spectral properties of GRBs on a statistical ground. In particular,
the characterisation of the time--averaged photon spectra is important
because it offers clues for understanding the radiation and particle acceleration
mechanism at work during the prompt phase of GRBs, on which there is no consensus yet.
In this paper, we extracted and studied the time--integrated photon spectra
of the 200 brightest GRBs observed with the Gamma--Ray Burst Monitor
(GRBM; \citealt{Feroci97,Frontera97}) aboard
the BeppoSAX mission \citep{Boella97} and performed a novel statistical study of the
main parameters characterising the GRB spectra.

The paper is organised as follows. Sections~\ref{sec:obs} and \ref{sec:an} report
the observations, data reduction, and analysis.
We report our results in Sect.~\ref{sec:res}, 
in the light of the models proposed in the literature,
and Sect.~\ref{sec:disc} presents our discussion and conclusions.

All quoted errors are given at 90\% confidence level for one interesting
parameter ($\Delta\chi^2=2.706$), unless stated otherwise.

\section{Observations}
\label{sec:obs}
The GRB sample used for this analysis was extracted from the GRB catalogue
of the BeppoSAX/GRBM (\citealt{Frontera09}, hereafter, F09).
The main constraints in the selection process were the following:
\begin{itemize}
\item sufficient number of total counts on the most illuminated detector unit;
\item well defined response function, connected with the information on the
GRB arrival direction;
\item reliable background interpolation.
\end{itemize}
In order to be able to derive a reliable time-integrated spectrum, we
noted that the average threshold on the number of total counts for a given
detector unit ranges from 3000 to 4000, depending on the GRB local direction to
BeppoSAX and on the GRBM unit considered in each case.

A number of GRBs (28) have also been detected in common with BATSE, 
whose data were published by K06 (see Sect.~\ref{sec:bep_bat}).
For these bursts, in addition to exploiting the information on the GRB position
derived by BATSE, useful to choose the appropriate response function, we compared
the results we obtained with the GRBM data with what published by K06.

The background interpolation and subtraction required the availability of spectra
acquired within contiguous time intervals around that/those including the burst.
This requirement further limited the final number of selected events.

Finally we ended up with 185 bright GRBs out of the 1082 GRBs belonging to the
GRBM catalogue (F09).
Hereafter, fluence $\Phi$ is referred to the 40--700~keV energy band, unless otherwise
specified.
The values of the largest and lowest fluences included in the final sample are
$1.7\times10^{-4}$ and $4.4\times10^{-6}$~erg~cm$^{-2}$, respectively.

\section{Data reduction and analysis}
\label{sec:an}
Firstly, among the 128-s time intervals continuously sampled with a time-integrated spectrum,
we identified those including the GRBs and those adjacent, required for the background estimate. 
In some cases we took only the most illuminated unit for each GRB; in the remaining cases,
we considered the two most illuminated units, apart from a few cases, for which it was possible
to extract meaningful spectra from three different units. The two-unit case typically occurred
when the burst direction with respect to the BeppoSAX local frame was such as to give comparable
counts to both units. Data from the second most illuminated unit were ignored when the
signal-to-noise (S/N) did not allow a statistically significant spectral reconstruction.

Table~\ref{tab:packets} reports the details of the data available for each analysed burst:
for each spectrum the corresponding time interval is referred to the on-ground trigger time (F09),
expressed as seconds of day (SOD). Each spectrum of a given burst is tagged with a letter
and the corresponding packet number inherited from the GRBM archival data (and used
in the GRB catalogue web interface\footnote{Available at \tt{http://saxgrbm.iasfbo.inaf.it}}) is also reported.

The GRB spectra reduction and analysis is performed as follows:
\begin{itemize}
\item dead-time correction of both source and background 128-s long spectra
(Sect.~\ref{sec:deadtime});
\item background fitting by interpolation of adjacent 128-s dead-time corrected
spectra (Sect.~\ref{sec:bkg});
\item identification of the appropriate GRBM response function (dependent on the GRB
position; Sect.~\ref{sec:rmf});
\item spectral fitting of background-subtracted GRB spectra:
  \begin{itemize}
  \item GRB total spectrum (``time-integrated'' spectrum);
  \item individual 128-s spectra of the GRBs that happened to be split into two
or more intervals (``time-resolved'' spectra).
  \end{itemize}
\end{itemize}

\subsection{Dead-time correction}
\label{sec:deadtime}
We set up the following procedure to correct the 128-s integrated spectra 
for dead time. The 40--700~keV light curve of the corresponding unit is taken into
account because a nearly constant rate gives rise to a dead time effect
smaller than a rate with prominent peaks of small time duration. For a given average 128~s 
spectrum, let $\tilde{c}_i$ be the observed counts in the 40--700~keV band for the $i$-th 1-s bin
($i=1,\ldots ,128$). 
The dead-time corrected counts, $c_i$ are calculated as $c_i = \tilde{c}_i/(1 - \tau\,\tilde{c}_i)$,
where $\tau=4$~$\mu$s is the dead time.
Let $f_i$ be the rate fraction of the corresponding bin, defined as $f_i = c_i/c$, where
$c$ is the sum of all $c_i$'s ($i=1,\ldots ,128$).
Let $\tilde{s}$ the total number of measured counts in the 128-s spectrum integrated over the 240
energy channels; this must also satisfy the following:

\begin{equation}
\tilde{s} = \sum_{i=1}^{128} \frac{f_i\ s}{1+\tau f_i\ s}
\end{equation}

where $s$ is the number of the total corrected counts we would have observed
integrating the spectrum in the absence of dead time. Therefore $s$ can be estimated as the root of
the following equation:

\begin{equation}
f(s) = \sum_{i=1}^{128} \frac{f_i s}{1+\tau f_i s} - \tilde{s} = 0.
\end{equation}

Assuming a negligible distortion of the original spectral shape due to dead time (which is the
case when no strong spectral evolution occurs during the 128-s interval over which the spectrum
is integrated), we renormalise the observed counts of each energy channel by the factor $s/\tilde{s}$.

\subsection{Background subtraction}
\label{sec:bkg}

In order to reliably estimate the background counts in each energy channel
of the time-integrated spectra, and to ensure a safe interpolation, we made
sure that spectra accumulated over 128-s time intervals temporally contiguous
to that including the burst were also available. In a few cases, either the time interval
preceding or that following the GRB was not available; in these
cases we checked that the background was so stable (typically within a few \%)
as to ensure linear (back)-extrapolation. For this reason we did not use the
128-s taken right before the ingress or right after the exit of a passage over
the South Atlantic Anomaly.

\begin{figure}
\centering
\includegraphics[width=9cm]{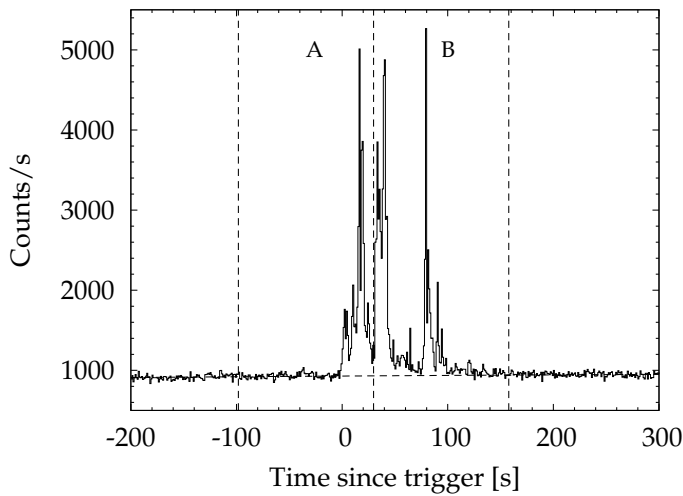}
\caption{1-s light curve of GRB~000226 detected with GRBM unit 2 in the
40--700~keV band shown as an example. The dashed horizontal line shows the
parabolic fit to the background. The vertical lines mark the time intervals
corresponding to the 128-s spectra continuously acquired. Spectra ``A'' and ``B''
include the GRB, while the adjacent intervals are used to interpolate the
background in each channel.}
\label{f:example_000226}
\end{figure}

Figure~\ref{f:example_000226} shows the example of GRB~000226
(\#~138 in Table~\ref{tab:packets}). The independent spectral sampling happened
to split this burst in two different 128-s spectra, called ``A'' and ``B'',
respectively (marked by vertical lines).
The light curve shown is the 40--700~keV profile of the most
illuminated unit (GRBM~2): the background was interpolated with a parabolic
fit (as in general), although a linear fit was already satisfactory.
%
%
We identified the packet numbers corresponding to ``A'' and ``B'':
30 and 31, respectively. We also took two couples of adjacent spectra
preceding (28, 29) and following (32, 33) the burst spectra, respectively.
All these spectra were dead-time corrected as in Sect.~\ref{sec:deadtime}.

\begin{figure}
\centering
\includegraphics[width=8.5cm]{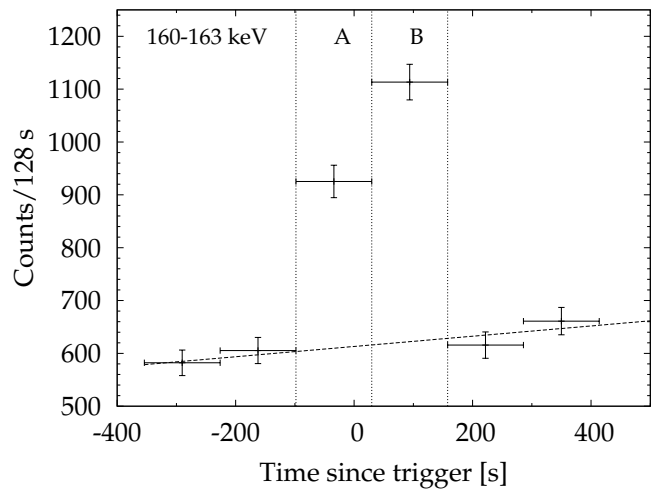}
\caption{128-s light curve of energy channel 60 (160--163~keV) of GRB~000226
as seen with GRBM unit 2 (same example as in Fig.~\ref{f:example_000226}).
The dashed line shows the linear fit of the background. Same vertical lines
as in Fig.~\ref{f:example_000226}.}
\label{f:example_000226_channel}
\end{figure}

At this point we performed two operations closely connected with one another:
background fitting and energy channels' grouping. The latter operation is the result
of ensuring a minimum significance (3$\sigma$) on the net counts of the final
background-subtracted grouped energy channel.
More in detail, this is how the procedure works: it extracts the light curve of a given
(original) energy channel out of the selected spectra. This light curve is then fit
linearly (excluding the spectra including the GRB) so that the interpolated background counts
expected during the GRB intervals are estimated. 
These steps are repeated for a sequence
of adjacent energy channels: every time an original energy channel is summed up,
the light curve extraction and fitting is reiterated for the grouped energy channel,
until the total net counts exceed the significance threshold.
When this is the case, the used energy channels are grouped into a final single channel.
An example of this is displayed in Fig.~\ref{f:example_000226_channel} for the same
burst shown in Fig.~\ref{f:example_000226}: the light curve of energy channel 60
(in this case corresponding to the energy range 160--163~keV) is built up from 6
contiguous 128-s time intervals. A satisfactory linear fit is performed on the
background intervals (dashed line): the resulting net counts in the GRB intervals
(``A'' and ``B'') match the significance requirement so that this channel can stand
alone.
%
%

The above steps are repeated until the full range of energy channels is covered.
The last grouped channel, that in general does not fulfil the significance requirement,
is merged into the previous one.
The goodness of the fit of the background light curve for each grouped energy channel
is expressed in terms of reduced $\chi^2$. This is then checked by the human operator
to make sure that all the fits are acceptable.
The uncertainty on the net counts of each grouped channel is calculated by propagating
the statistical uncertainties of counts and those affecting the interpolated background.

All this applies to the meaningful channels, i.e. from 18 to 240 out of the 256 nominal
channels, while the remaining are ignored.

\subsection{Response Matrices}
\label{sec:rmf}
The knowledge of the appropriate response matrices for a given burst requires
the GRB position with respect to the BeppoSAX reference frame to be known, also
called ``local'' position. The reason is the complex dependence of
the response function on the local direction and photon energy, due to the
BeppoSAX payload itself surrounding the GRBM units.
The GRBM response function for a generic direction was determined with Monte
Carlo techniques \citep{Calura00} and in-flight calibrated with Crab observations
and cross-calibrated with BATSE through commonly detected GRBs.
See F09 for a detailed description.

The information on the directions of the GRBs considered in this work was taken
from F09: column ``CAT'' in Table~\ref{tab:packets} reports the ID
of the catalogue providing the most accurate position of each GRB using the same
convention as in F09. For those GRBs for which no such information is available,
typically the GRBs detected by the GRBM alone and for which the localisation
procedure did not give a unique acceptable solution, we used as many response
matrices as the possible directions and made sure that the spectral results
were not significantly different from each other.

\subsection{GRB spectra and models}
\label{sec:grb_spec}
In the case of GRBs whose profile was sampled by multiple 128-s intervals, we
extracted and fit both the total (time-integrated) and the individual (time-resolved)
spectra. Whenever the burst was contained within a single interval, no
time-resolved spectrum was possible.

We adopted three possible fitting models: i) Band's model ({\sc band}; \citealt{Band93})
ii) the cut-off power--law ({\sc cpl}), where the photon spectrum is
$N(E)\propto E^{-\alpha}\,\exp{[-E\,(2-\alpha)/E_{\rm p}}]$;
iii) a simple power--law ({\sc pow}), $N(E)\propto E^{-\alpha}$.

In addition to the normalisation, the free parameters of the fitting models were
the power--law indices (the low- and high-energy $\alpha$ and $\beta$ for {\sc band},
only the low-energy index for the other models) and the peak energy, $E_{\rm p}$, of
the $\nu$$F_\nu$ spectrum. We note that the signs of the {\sc band} photon indices
follow a different convention from the other models.
Spectral fitting was done using XSPEC v12.5 \citep{Arnaud96}.

Table~\ref{tab:results} reports the spectral fitting results for the total spectra
of all the bursts for various models. The {\sc pow} model was adopted when the goodness
of the fit, expressed through the reduced $\chi^2$, was already acceptable and 
fitting with the other models did not provide any useful constraint on $E_{\rm p}$.
This was typically the case for bursts with the $E_{\rm p}$ either above or below
the energy passband of the GRBM and/or for spectra with relatively poor S/N.
Following \citet{Sakamoto08a}, for each of the 185 time-integrated spectra we first fit each
spectrum with each of the three models. Whenever passing from a model to a more complex
one the total $\chi^2$ decreased by more than 6 for each additional degree of freedom,
we considered it a significant improvement in modelling the spectrum (see \citealt{Sakamoto08a}).

In most cases, the high-energy photon index $\beta$ of the {\sc band} function could not be
constrained by the data, due to the narrower energy passband of the GRBM compared with that of
BATSE as well as to the S/N ratio of the spectra. 
In such cases, we fixed its value to the average value of $-2.3$ found on the BATSE sample
(K06). In a few cases the same problem occurred at low energies, for which we fixed the
corresponding index $\alpha$ to the analogous value of $-1.0$.

We have not achieved in any case a statistical improvement with the {\sc band} model,
except for three cases of time-resolved spectra, characterised by a large S/N
(spectra B of 980615B,  971208B and 970831). The best--fit model of each GRB is marked with an asterisk
in Tables~\ref{tab:results} and \ref{tab:results_time}.

We excluded from statistical analysis the bursts whose spectra gave a poor fit,
i.e. whose results in terms of $\chi^2/{\rm dof}$ can be rejected at 99\% confidence level.
These cases represent less than 3\% of the total sample and are the following bursts:  
980203B, 990118A, 000328, 001213 and 001228.

\section{Results}
\label{sec:res}

\subsection{Results in the BeppoSAX local frame}
\label{sec:local_sax}
We studied the goodness of the spectral fit for each GRB as a function
of the direction as referred to the BeppoSAX local frame of reference (F09).
The aim is to check the goodness of the response matrix as a function
of the GRB arrival direction. To this aim, we investigated how the total
reduced $\chi^2$ for each GRB best-fitting spectral model depends on both
the local azimuth $\phi$, measured counterclockwise from the axis of
GRBM unit 2, and the local altitude $\theta$ above the BeppoSAX equatorial
plane, respectively.
%
\begin{figure}
\centering
\includegraphics[width=9cm]{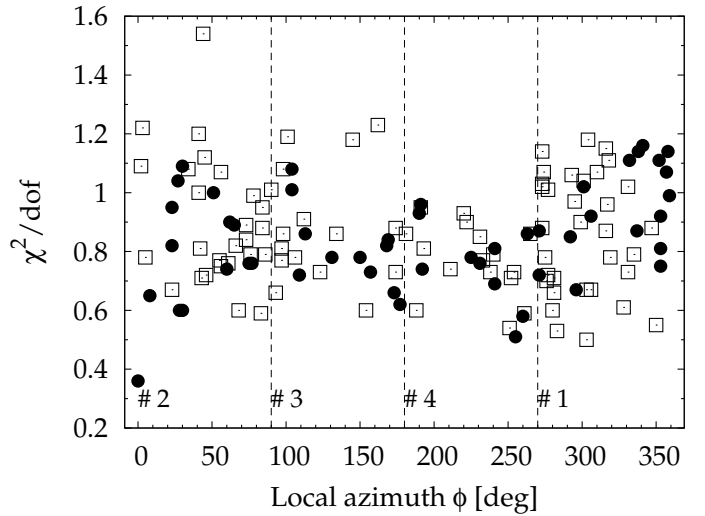}
\caption{Goodness of the fit for all the brightest GRBs with known arrival
direction (143 out of 200) as a function of the BeppoSAX local azimuth angle
$\phi$. Vertical dashed lines at $\phi=0^\circ, 90^\circ, 180^\circ, 270^\circ$ correspond to
the axes of GRBM units 2, 3, 4, 1, respectively. Empty squares (filled circles)
are the GRBs localised by other experiments (GRBM alone), as reported in F09.}
\label{f:chi_vs_phi}
\end{figure}
%
%
\begin{figure}
\centering
\includegraphics[width=9cm]{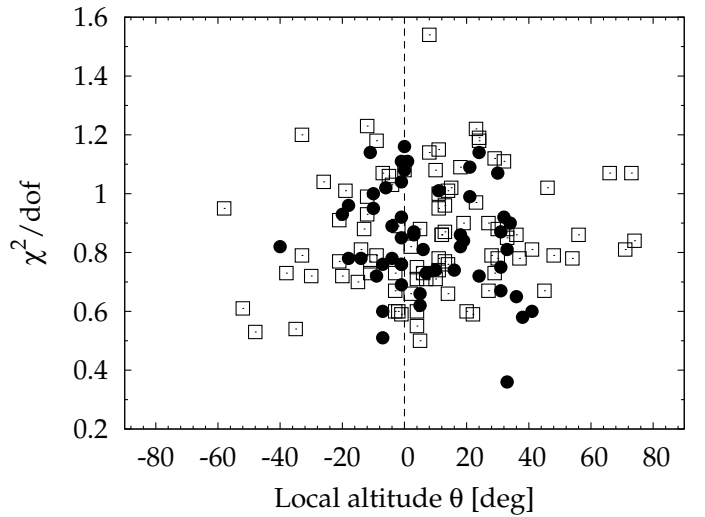}
\caption{Same as Fig.~\ref{f:chi_vs_phi}. Angle $\theta$ is
the BeppoSAX local altitude above the equatorial plane,
marked by the vertical dashed line.}
\label{f:chi_vs_the}
\end{figure}
%

Figure~\ref{f:chi_vs_phi} shows the reduced $\chi^2$ of the best-fitting spectral
model for each GRB as a function of the local azimuthal angle $\phi$ for 143 GRBs
with known arrival direction out of our sample. The points are divided in two classes:
those localised by other experiments (empty squares), whose information on the
direction is independent from the GRBM, and the remaining ones localised with the
GRBM (filled circles; see F09).
Clearly both classes do not show any strong dependence of the goodness of the fit
on the local azimuthal direction. We just note that GRBs close to GRBM unit 2 axis
have slightly more scattered $\chi^2$ values than other units.

Figure~\ref{f:chi_vs_the} shows the goodness of the fit as a function of
the local elevation or altitude angle $\theta$. As for the azimuthal angle,
the goodness of the fit shows no dependence on the elevation angle either.
Here noteworthy is the presence of more GRBs (60\% of the total) in the BeppoSAX northern
hemisphere ($\theta>0^\circ$):
this is explained by the more effective absorption for southern directions
due to the on-board electronics boxes in the lower part of the spacecraft
\citep{Guidorzi02}, as evidenced by the number of bright GRBs, that
drops significantly for elevation angles below $-20^\circ$/$-30^\circ$.
The paucity of GRBM--localised GRBs compared with those
localised by other instruments at directions close to the BeppoSAX local poles
is due to the limitations of the GRBM localisation technique (F09).

\subsection{Results of the {\sc pow} model}
\label{sec:res_pow}
The power--law model provides acceptable fits for $\sim 35$\% of the
sample. This model represents the best-fitting model
for 10\% of the 100 brightest GRBs; the same fraction rises to 53\%
when we consider the less bright half of the sample.

The power--law index $\alpha_{\rm pow}$ distribution was derived by selecting only
those GRBs whose spectral fitting gave an uncertainty smaller than $0.3$.
This choice was the result of a trade-off between the need of reasonably
accurate values and the need of a good statistics.
As a consequence the sample shrank to 87\%.

\begin{figure}
\centering
\includegraphics[width=9cm]{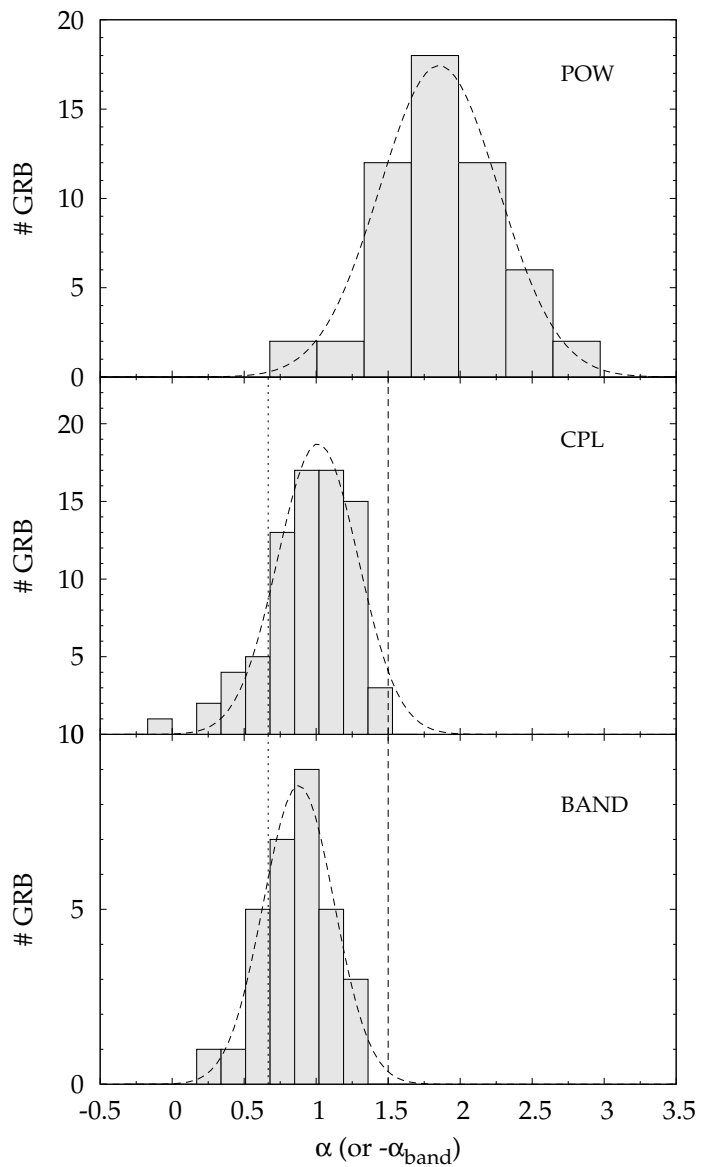}
\caption{ {\em Top panel}: $\alpha_{\rm pow}$ distribution for 55 GRBs with
relative uncertainties smaller than 0.3.
{\em Mid panel}: $\alpha_{\rm cpl}$ distribution for 77 GRBs with relative
uncertainties smaller than 0.5.
{\em Bottom panel}: alternatively to the {\sc cpl} model, we show the
$-\alpha_{\rm band}$ distribution for 31 GRBs for which the {\sc band} function
provides an acceptable fit, although not significantly better than the {\sc cpl}.
The vertical dotted and dashed lines show the cases $\alpha=2/3$
(synchrotron death line) and $\alpha=3/2$ (cooling death line).
In each panel dashed distributions show the corresponding
best-fitting Gaussian functions.}
\label{f:alpha_hist}
\end{figure}

The resulting distribution of $\alpha_{\rm pow}$ can be fit with a Gaussian with
$\overline{\alpha}_{\rm pow}=1.86$ and $\sigma(\alpha_{\rm pow})=0.32$ (top panel
of Fig.~\ref{f:alpha_hist}),
in agreement within the analogous results obtained over a sample by
Swift/BAT ($1.6 \pm 0.2$; \citealt{Sakamoto08a}) as well over a sample of
INTEGRAL ($1.6$ in the energy band 18--300 keV; \citealt{Vianello09}). 
The slightly softer average value we obtained with the GRBM bursts is explained
by the harder energy range considered, thus more likely to be affected by the
steepening of the spectrum due to the high-energy component; this conclusion
is also supported by the corresponding value ($\overline{\alpha}\simeq 1.7$)
obtained over the BATSE sample (K06).

The softness of most spectra fit with a {\sc pow} model is explained
with most GRBs having $E_{\rm p}$  below $\sim 100$~keV. 
About 30\% of the {\sc pow} indices lie in the range $1.7<\alpha_{\rm pow}<2.0$,
so their peak energies are likely to lie close to the upper bound or above it,
i.e. $E_{\rm p}\gtrsim700$~keV.
Another 37\% of the same sample have $\alpha_{\rm pow}>2$ and their peak energies
lie at $E_{\rm p}<40$~keV, as expected for the X--Ray Flashes (XRFs; \citealt{Heise01,
Barraud03,Sakamoto05,Sakamoto08b,Pelangeon08}).

\subsection{Results of the {\sc cpl} model}
\label{sec:res_cpl}
Similarly to the {\sc pow} model case, the distribution of the power--law index
for the {\sc cpl} model was derived selecting the GRBs with an uncertainty
on $\alpha_{\rm cpl}$ smaller than 0.5. 77\% of the sample passed this criterion.

The resulting distribution is fit with a Gaussian with mean and standard
deviation values of $\overline{\alpha}_{\rm cpl}=1.0$ and
$\sigma(\alpha_{\rm cpl})= 0.28$ (mid panel of Fig.~\ref{f:alpha_hist}).

Similar values were obtained from the observations of 
BAT aboard Swift: $\alpha_{\rm cpl} = 1.12 \pm 0.15$  \citep{Cabrera07},
HETE-II: $\alpha_{\rm cpl} = 1.2 \pm 0.5$ \citep{Barraud03}.
There are no cases in which the low--energy index is very soft
($\alpha\ge 2$).

In the past, spectral fitting of time--resolved BATSE spectra of bright GRBs
has yielded a significant number of cases with low--energy photon indices
$\alpha$ below 2/3. This result is inconsistent with the SSM and $\alpha=2/3$
has been referred to as its ``death line''
(\citealt{Preece98}, \citealt{Papathanassiou99} and references therein).
Our results indicate that about 30\% of them lie below the synchrotron death line
of $\alpha=2/3$ (vertical dotted line in Fig.~\ref{f:alpha_hist}).
On the other side of the distribution, no GRB lie beyond
the fast-cooling death line \citep{Ghisellini00} represented by the limit
$\alpha<3/2$ (vertical dashed line).

\begin{figure}
\centering
\includegraphics[width=9cm]{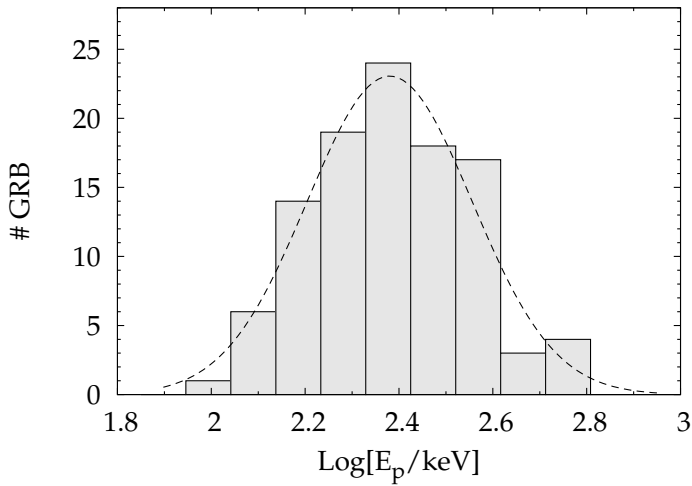}
\caption{Peak energy distribution derived with the {\sc cpl} model for
a sample of 106 GRBs with a relative uncertainty smaller than 40\%.
The dashed line shows the best-fitting normal distribution.}
\label{f:peak_energy}
\end{figure}

Figure~\ref{f:peak_energy} shows the $E_{\rm p}$ distribution for the {\sc cpl}
model. Only values with uncertainties smaller than 40\% are displayed; they
represent 90\% of the overall set of GRBs best fit with {\sc cpl}.
The distribution can be fit with a log--normal with mean and standard deviation
of $\log{E_{\rm p}} = 2.38 \pm 0.18$ (corresponding to a mode
of $E_{\rm p}=240$~keV), fully consistent with the results obtained over a sample of bright BATSE
bursts by K06: they found $E_{\rm p} = 251^{+122}_{-68}$~keV using different fitting models.

%
%

Comparing the $E_{\rm p}$ distribution of our sample with the analogous of the BATSE GRBs
fit only with the {\sc cpl} model ($E_{\rm p} = 321^{+202}_{-105}$~keV), although formally
consistent with one another, suggests that the GRBM distribution is shifted towards lower
values: this is primarily explained by the BATSE sensitivity at energies $> 700$~keV.
The consequence of this selection effect is that a number of bright GRBs
detected with the GRBM and with $E_{\rm p}\gtrsim700$~keV are clearly missing
in the observed $E_{\rm p}$ distribution of Fig.~\ref{f:peak_energy}, and belong
to the GRBs that were fit with a power--law with $\alpha_{\rm pow}<2$.

We do not observe a sizable fraction of GRBs with $E_{\rm p}<100$~keV because 
our sample collects the brightest end of the GRBM fluence distribution:
as a consequence, our sample is biased towards GRBs with high $E_{\rm p}$ values,
because of its correlation with the fluence (Fig.~\ref{f:ep_vs_flu}).

\begin{figure}
\centering
\includegraphics[width=9cm]{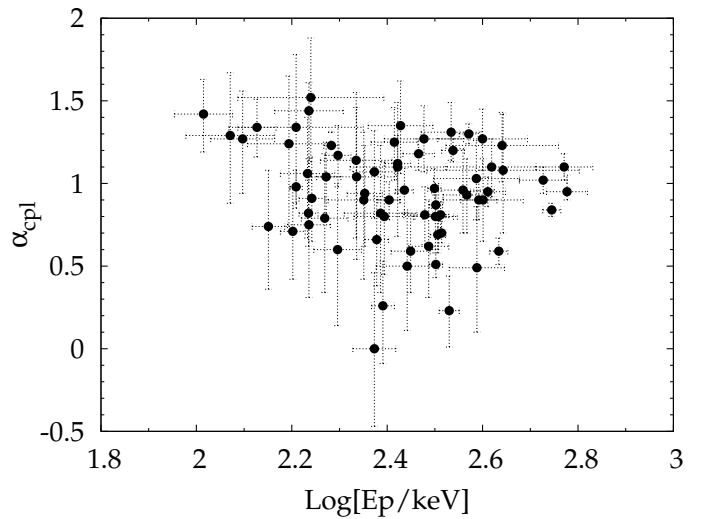}
\caption{Low--energy photon index vs. peak energy as determined with the
{\sc cpl} model on a sample of 70~GRBs with relatively accurate measurements.}
\label{f:alpha_vs_logEp}
\end{figure}

We analysed the possible relation (if any) between $\alpha_{\rm cpl}$ and $E_{\rm p}$
for a sample of 70~GRBs with both measurements sufficiently accurate by adopting
the same thresholds mentioned above ($0.5$ on $\alpha_{\rm cpl}$ and 40\% on $E_{\rm p}$).
Figure~\ref{f:alpha_vs_logEp} shows this sample in the $\alpha_{\rm cpl}$--$E_{\rm p}$ plane.
%
%
A statistical study of our data shows no clear evidence for a correlation
between these two quantities. In fact,
the Spearman rank--order correlation coefficient over the whole sample turned out
to be $r_s=-0.15$ with an associated probability of 21\% of no correlation.
However, for the low peak energy subsample ($\log{E_{\rm p}/{\rm keV}}<2.4$) the
probability drops to $0.6$\% ($r_s=-0.49$), suggesting that the softer
the peak energy, the softer the photon index.
This type of correlation is expected because of the instrumental
effect coming into play whenever $E_{\rm p}$ lies close to the edge of the energy
passband. When this is the case, the low--energy photon index $\alpha$ as derived
from the fitting procedure may have not reached the asymptotic value, thus
resulting in a softer value \citep{Preece98,Lloyd00,Lloyd02,Amati02}.
This is indeed what we observe in Fig.~\ref{f:alpha_vs_logEp}.
That such a correlation seems to become significant when considering only the
GRBs with low $E_{\rm p}$ values is a clear indication of its instrumental origin.

The number of GRBs whose $\alpha$ estimates are biased because of this effect
depends on how smoothly the spectrum reaches its asymptotic value, in addition
to the energy window of the detector. The same problem also affects samples of GRB
spectra obtained with different detectors. To circumvent this issue, in the case
of BATSE GRBs \citet{Preece98} defined the ``effective low--energy photon index'' as the
tangential slope at 25~keV (lower energy bound of BATSE detectors) of the spectrum
in logarithmic scale. However, also with this definition the problem still remains
whenever 25~keV is not low enough to reach the asymptote \citep{Lloyd00}.
In our case, the impact on the $\alpha$ distribution shown in Fig.~\ref{f:alpha_hist} is
such that a few GRBs with $1.3\lesssim\alpha_{\rm cpl}\lesssim1.5$ are likely to suffer from this
effect. Different detectors with different energy windows should be also affected
differently. However, as we noted above, the analogous distributions of other detectors,
described by similar modes and dispersions, suggest that the impact of this instrumental effect
on the observed $\alpha$ distribution is minimal.

\subsection{Results of the {\sc band} model}
\label{sec:res_band}
In none of the time--integrated spectra of our sample we found a significant improvement
when changing the fitting model from {\sc cpl} to {\sc band}.
Nevertheless, to explore how the choice of either model may affect the result, in
the bottom panel of Fig.~\ref{f:alpha_hist} we show the $-\alpha_{\rm band}$ distribution
for 31 GRBs for which the {\sc band} function gave an acceptable result, although not
significantly better than the {\sc cpl}. Clearly, the two distributions are fully
compatible with each other.
In Section~\ref{sec:cpl_vs_band} we explore in more details the relation between
the usage of the two models with our data.

\subsubsection{High--energy index}
\label{sec:band_beta}
\begin{figure}
\centering
\includegraphics[width=9cm]{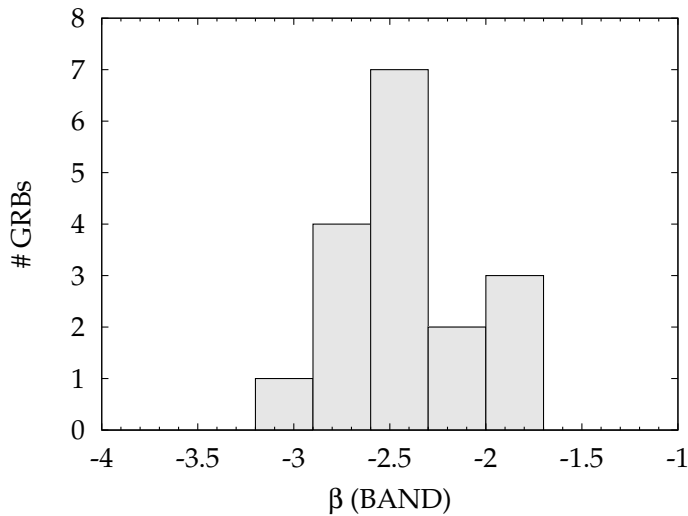}
\caption{High--energy photon index $\beta$ distribution of
the {\sc band} function for a sample of 17 GRBs.}
\label{f:beta_free_hist}
\end{figure}

For a subsample of 17~GRBs out of the 31 mentioned above, it was also possible to
constrain the high--energy photon index $\beta$ with absolute uncertainties
smaller than 1.
The distribution is shown in Fig.~\ref{f:beta_free_hist}.
%
%
The small number of events for which this estimate was possible is explained
by the relatively small upper bound of the GRBM passband compared with
that of BATSE. However, we note that the resulting distribution is
fully compatible with that derived on a more numerous BATSE sample (K06).

\subsection{Fluence distribution}
\label{sec:fluence}
To account for the uncertainties in the response matrix calibration, we added in quadrature
10\% systematic to the statistical fluence errors (F09). For each GRB we considered the fluence
yielded by the corresponding best-fitting model.
For the analysis we considered the mean value and symmetric error of the corresponding logarithms.
The distributions of the best-fitting parameters and of the fluence were derived by
excluding the GRBs affected by a relative uncertainty larger than 20\% (after including
the systematics); 12\% of the sample were rejected as a consequence.

\begin{figure}
\centering
\includegraphics[width=9cm]{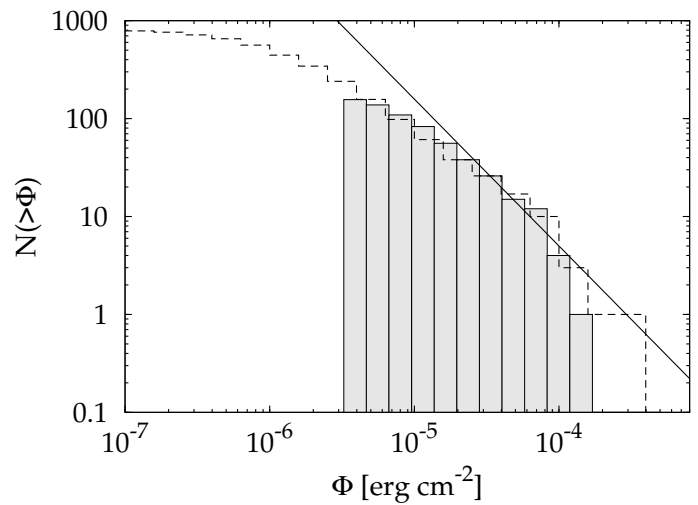}
\caption{Cumulative fluence distribution for the subset of GRBs with
a relative error on fluence smaller than 20\% (shaded histogram).
The dashed histogram is the corresponding fluence distribution published
by F09 for the entire GRBM catalogue of GRBs, as derived from 2--channel
spectra. The solid line shows the power--law distribution with index $-3/2$
expected if GRBs were homogeneously distributed in an Euclidean space
throughout the sampled volume.}
\label{f:fluence_distribution}
\end{figure}

Figure~\ref{f:fluence_distribution} displays the cumulative fluence distribution
(shaded histogram) compared with the corresponding distribution for 795 GRBs of the GRBM
catalogue by F09, whose values were calculated through the 2-energy channel spectra
(dashed histogram; see fig.~8 of F09). The solid line shows the power--law distribution
with index $-3/2$, predicted in the case of no luminosity function evolution with redshift and
where the observed GRBs are homogeneously distributed in the sampled volume of an
Euclidean space. The difference between the observed and the predicted $-3/2$ power--law 
distributions had already been found in the BATSE catalogue \citep{Meegan92} and
confirmed with the GRBM data (F09).
%

\subsection{Peak energy--fluence correlation}
\label{sec:ep_flu_corr}
\begin{figure}
\centering
\includegraphics[width=9cm]{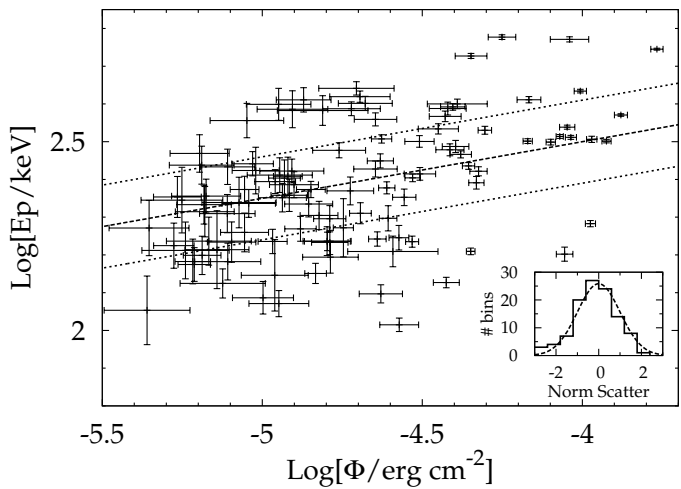}
\caption{Correlation between $E_{\rm p}$ and the 40--700~keV fluence
for a sample of 108 bright GRBs with well determined values.
The dashed line shows the best-fitting power--law, with index $0.21\pm0.06$,
while the dotted lines include the 1-$\sigma$ region, where $\sigma=0.13$
is the extrinsic scatter. {\em Inset}: distribution of the normalised
scatter. The dashed line shows the standardised normal distribution.}
\label{f:ep_vs_flu}
\end{figure}

Figure~\ref{f:ep_vs_flu} displays the observed peak energy $E_{\rm p}$ vs.
fluence $\Phi$ for a sample of 108 bright GRBs with well determined 
values. The correlation is significant: the Spearman rank coefficient is
$r_s=0.48$ with an associated chance probability of $1.4\times10^{-7}$.
Given the apparent scatter, we performed a power--law fit adopting the
D'Agostini method (e.g., \citealt{Guidorzi06}) and found the following
best-fitting relation:
\begin{equation}
\log{\Big(\frac{E_{\rm p}}{{\rm keV}}\Big)}\ =\ (0.21 \pm 0.06)\ 
\log{\Big(\frac{\Phi}{{\rm erg\ cm}^{-2}}\Big)}\ +\ (3.4 \pm 0.3)
\label{eq:ep_vs_flu}
\end{equation}
The extrinsic scatter, which combines and must not be confused with the intrinsic scatter
due to the uncertainties of the individual points, is $\sigma_{\log{E_{\rm p}}}=0.13\pm0.02$.
The slope of the correlation agrees with previous values obtained
on samples of BATSE GRBs (e.g. \citealt{Lloyd00b,Nava08}).
%
%
The scatter of each point in the $E_{\rm p}$--$\Phi$ plane around the best-fit
correlation is the result of the two sources of scatter: the intrinsic, different
for each point and accounting for the uncertainties in the evaluation process
of both observables, and the extrinsic scatter, reflecting a property of the
correlation itself through some unknown variables. The combination of the
two scatter sources finally give a normal distribution, as shown by
the inset of Fig.~\ref{f:ep_vs_flu}: indeed the normalised scatter $\zeta_i$, defined by
eq.~(\ref{eq:norm_scatter}) with $i$ running over the set of points, distributes
according to a standardised Gaussian.

\begin{equation}
\zeta_i\ =\ \frac{y_i - (m\,x_i + q)}{\sqrt{\sigma_{y,i}^2 + m^2\,\sigma_{x,i}^2 + \sigma_y^2}}
\label{eq:norm_scatter}
\end{equation}

where $y_i=\log{E_{{\rm p},i}}$, $x_i=\log{\Phi_i}$, the $i$-th $\sigma$'s being the
corresponding (intrinsic) uncertainties and $\sigma_y$ being the extrinsic one.
$m$ and $q$ are the best-fit slope and constant values reported in eq.~(\ref{eq:ep_vs_flu}),
respectively.

It is known that truncation effects connected with the finiteness of the detector energy
window may affect the distribution of the fitting parameters and the corresponding
correlations. In particular, both $E_{\rm p}$ and fluence $\Phi$ suffer from them, as
proven for BATSE GRBs by \citet{Lloyd00b}, who investigated their impact in this respect.
Our sample includes bright bursts, so truncation effects against low--fluence GRBs or
near the detector threshold can be neglected. As discussed in Sect.~\ref{sec:res_cpl},
$E_{\rm p}$ is more likely to suffer from biases against low and high values \citep{Lloyd99}.
\citet{Lloyd00b} tackled the issue of accounting for data truncation effects in the
correlation studies by means of non--parametric techniques, which take into account the
limits imposed by the detector in the determination of each parameter.
Specifically for the $E_{\rm p}$--$\Phi$ relation, they found similar results with these
techniques and when considering the bright subsample of BATSE bursts,
less affected than the GRBs close to the detector threshold. In both cases the value
of the best-fitting slope ($0.29\pm0.03$ and $0.28\pm0.04$, respectively) is similar
to that found on our set.
This suggests that the slope of the $E_{\rm p}$--$\Phi$ correlation for the bright end of the GRBs
detected with the GRBM is only marginally affected by this kind of truncation effects.

\subsection{{\sc cpl} vs. {\sc band}}
\label{sec:cpl_vs_band}
Although the {\sc band} provided a significant improvement in the spectrum fitting
only for a very few cases of time--resolved spectra (Sect.~\ref{sec:grb_spec}),
we studied how the estimate of a given parameter compares with that obtained with
the other model.
In this respect, we considered both the low--energy photon index and the peak
energy for a set of GRBs that could be fit with either model.

\subsubsection{Peak Energy}
\label{sec:cpl_vs_band_Ep}
\begin{figure}
\centering
\includegraphics[width=8cm]{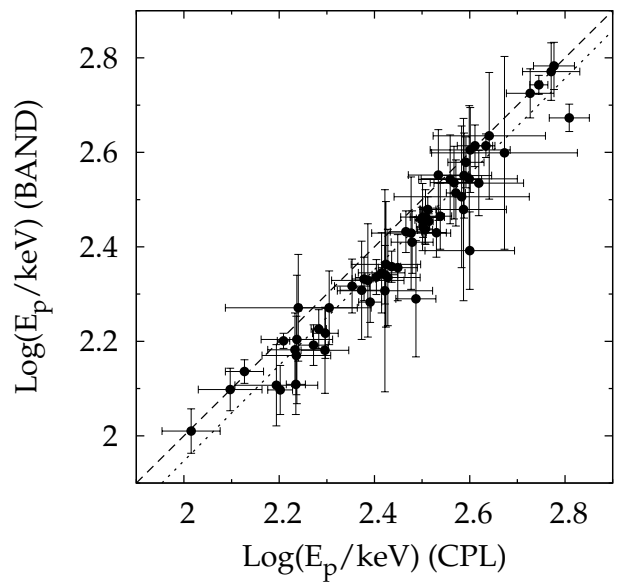}
\caption{Peak energy determined with the {\sc cpl} vs. the same determined
with the {\sc band} function for a sample of 63 GRBs with both measurements.
The dashed and dotted lines show the equality and the best-fitting power--law
relations, respectively. In particular, according to the latter, the peak energy
determined with the {\sc cpl} tends to be $\sim 20$\% larger than that of the
{\sc band}.}
\label{f:Ep_cpl_vs_band}
\end{figure}

Figure~\ref{f:Ep_cpl_vs_band} shows the comparison of $E_{\rm p}$ as determined with
the {\sc cpl} and that derived with the {\sc band} models for a sample of 63 GRBs
for which both models provided an acceptable result.
While for all GRBs they essentially provide consistent results within uncertainties,
a linear regression accounting for the uncertainties of all the points along both
axes shows that the {\sc cpl} model tends to slightly overestimate $E_{\rm p}$ by
$\sim 20$\% on average with respect to the {\sc band} function. This is proven
by the best-fit (dotted line in Fig.~\ref{f:Ep_cpl_vs_band}) described by
eq.~(\ref{eq:Ep_cpl_vs_band}).
\begin{equation}
\log{E_{\rm p,band}}\ =\ (1.012\pm 0.042)\ \log{E_{\rm p,cpl}} - (0.077\pm 0.102)
\label{eq:Ep_cpl_vs_band}
\end{equation}
However, within the level of accuracy of our data, the two models provide
equivalent estimates of $E_{\rm p}$ within uncertainties, as shown by the
best-fitting parameters in eq.~(\ref{eq:Ep_cpl_vs_band}),
fully consistent with equality.

%
%

\subsubsection{Low--energy index}
\label{sec:cpl_vs_band_alpha}
\begin{figure}
\centering
\includegraphics[width=8cm]{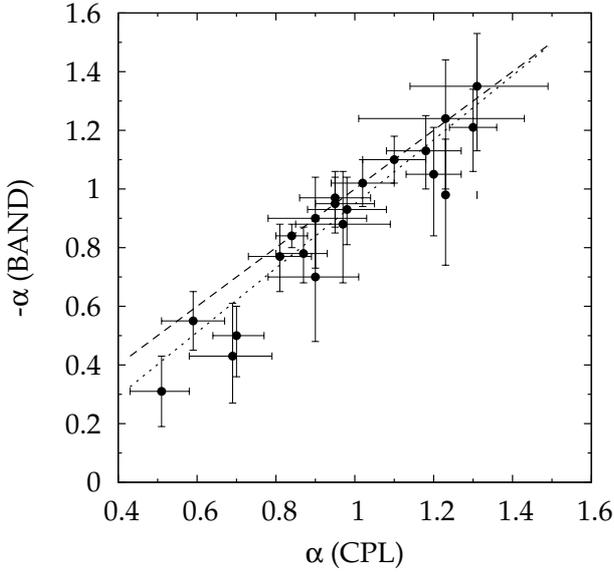}
\caption{Low--energy photon index determined with the {\sc cpl} vs. the same
determined with the {\sc band} function for a sample of 21 GRBs with both measurements.
The dashed and dotted lines show the equality and the best-fitting linear
relations, respectively.}
\label{f:alpha_cpl_vs_band}
\end{figure}

We selected a sample of GRBs with the low--energy photon index determined
from the spectral fitting with both models and required both uncertainties
to be smaller than $0.5$. In this way 21 GRBs were selected, shown
in Figure~\ref{f:alpha_cpl_vs_band}. For the sake of clarity, we consider
$-\alpha_{\rm band}$ to be compared with $\alpha_{\rm cpl}$.
The two models clearly provide consistent estimates for the low--energy
photon index, as shown by the equality line (dashed). Performing a linear
fit between the two sets taking into account the uncertainties along both
axes, the result is described by eq.~(\ref{eq:alpha_cpl_vs_band}) and
shown with dotted line in Fig.~\ref{f:alpha_cpl_vs_band}.
\begin{equation}
-\alpha_{\rm band}\ =\ (1.094\pm 0.152)\ \alpha_{\rm cpl} - (0.145\pm 0.146)
\label{eq:alpha_cpl_vs_band}
\end{equation}
The slightly lower values of $|\alpha_{\rm band}|$ are not statistically
significant, so in our sample we may consider the two models equivalent
as for the low--energy photon index estimate.
%
%

\subsection{GRBM vs. BATSE}
\label{sec:bep_bat}

In our sample there are 28 bursts observed also by BATSE whose spectral
fitting results were published by K06; they are marked in Table~\ref{tab:results}.

\begin{figure}
\centering
\includegraphics[width=7.5cm]{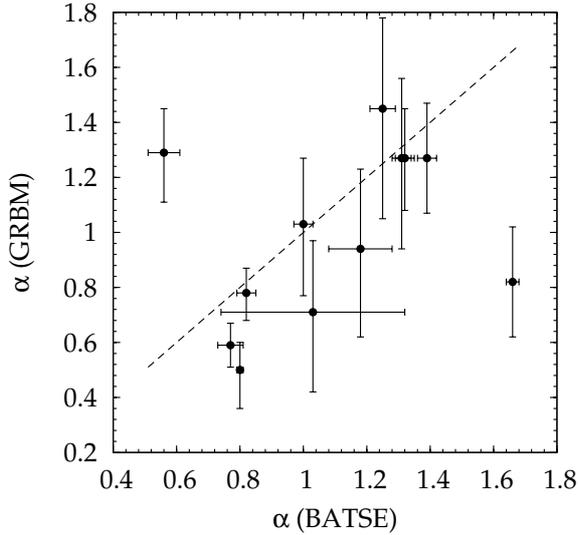}
\caption{BeppoSAX/GRBM versus BATSE: low-energy power--law spectra
as measured with {\sc band} and {\sc cpl} models.
The dashed line shows the equality line.}
\label{f:BEPvsBAT_alpha}
\end{figure}

For the sake of homogeneity, we considered the low--energy photon index values of BATSE
obtained from fitting instead of the so--called effective values \citep{Preece98} estimated by K06.
As shown in Fig.~\ref{f:BEPvsBAT_alpha}, the values of $\alpha$ of the GRBM
sample look harder than BATSE, but this is not really significant and is merely due to
the larger uncertainties of the former. There are a couple of cases with significantly
different values for the two instruments: 970831 ($4.2\sigma$) and 971220 ($4.4\sigma$).
We investigated the possible reasons for this discrepancy.
In the case of 970831 K06's $\alpha$ is much softer than ours and this could be due
to the different time intervals used: K06's spectrum missed the first $\sim 20$~s
of the $\simeq 150$ s long burst. As for 971220, K06 integrated from BATSE trigger
out to $9.5$~s, whereas the burst lasted at least up to $\sim 15$~s, missing
the last part of the profile; the GRBM spectrum could be fit with a single
power--law with $\alpha=1.3\pm0.2$, consistent with the BATSE $E_{\rm p}$ of $\sim 2$~MeV.
However, their $\alpha$ estimate, taken from the {\sc cpl} model, is reported to
be $\alpha=0.56\pm0.05$, i.e. much harder. Using the {\sc cpl} by K06 might contribute
to give a different value for $\alpha$ from that of the {\sc pow} obtained by us; however this is due
to the intrinsic curvature of the {\sc cpl}.
%
%

\begin{figure}
\centering
\includegraphics[width=7.5cm]{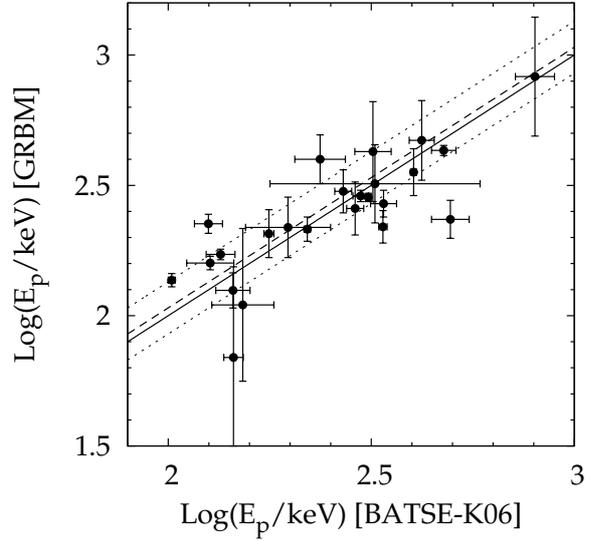}
\caption{BeppoSAX/GRBM versus BATSE: peak energy
as measured with {\sc band} and {\sc cpl} models.
The solid line shows the equality line. The dashed
and dotted lines show the best-fitting relation 
and the 1--$\sigma$ region.}
\label{f:BEPvsBAT_logEp}
\end{figure}

Figure~\ref{f:BEPvsBAT_logEp} displays the values of the peak energy
for the sample of 28 common GRBs as determined from the two data sets.
The points scatter around the equality line (solid):
we quantified such scatter applying the D'Agostini method
by fixing the slope of the relation to 1 and leaving the constant term as
well as the extrinsic scatter free to vary.
%
%
The best-fitting result is shown by the dashed line, whereas the dotted
lines identify the 1--$\sigma$ region around the best fit.
Equation~(\ref{eq:BEPvsBAT_logEp}) describes the best-fitting function:
\begin{equation}
\log{\Big(\frac{E_{\rm p,GRBM}}{{\rm keV}}\Big)}\ =\ m\,\log{\Big(\frac{E_{\rm p,BATSE}}{{\rm keV}}\Big)}\ +\ q
\label{eq:BEPvsBAT_logEp}
\end{equation}
The best-fitting values are $m=1$ (fixed), $q=0.03\pm0.05$,
$\sigma(\log{E_{\rm p,GRBM}})=0.10^{+0.07}_{-0.04}$. The origin of this
scatter, corresponding to $\sim 26$\%, which adds in quadrature to
the uncertainties of the individual points, must be searched in a combination
of factors: i) the different integration time intervals, whose choice is forced
by the different spectral sampling of the light curves of the two
instruments; ii) the different energy passband; iii) different geometry
GRB direction--Earth--instrument aboard the two spacecraft (with
different albedo effects).

In practice, we note that an additional uncertainty of 26\% in the
time-average peak energy $E_{\rm p}$ estimate does not affect appreciably
any correlation between $E_{\rm p}$ and other relevant observables,
such as the $E_{\rm p,i}$--$E_{\rm iso}$ relationship \citep{Amati02}.

When we release the $m=1$ constraint, we found a significant shallower
dependence of $E_{\rm p,GRBM}$ on $E_{\rm p,BATSE}$, $m=0.61\pm0.14$, not
shown in Fig.~\ref{f:BEPvsBAT_logEp}. This is due to data truncation,
as discussed in Sect.~\ref{sec:ep_flu_corr}, and is explained by the narrower
passband of the GRBM with respect to that of BATSE: the former tends
to move inside the 40--700~keV range those values of $E_{\rm p}$ whose
BATSE measurements, thanks to its broader energy range especially at
high energies, are likely to be less biased by the finite energy range.

Even for $E_{\rm p}$ there are a few GRBs with significantly different
values: 970616 ($5.0\sigma$), 971029 ($4.8\sigma$) and 990718 ($3.8\sigma$).
The case of 970616 is peculiar, since it occurred when the BeppoSAX
spacecraft was temporarily unstable due to the loss of gyroscopes in
May--June 1997. As a consequence, the BeppoSAX local direction of this
BATSE burst is not known and, even worse, throughout the duration of 
the burst (lasted about 80~s) the pointing was not constant. All this
might have determined the discrepancy in the measurement of $E_{\rm p}$,
which is enhanced by the smallness of the uncertainties provided by K06:
$E_{\rm p,BATSE}=(102\pm2)$~keV to be compared with our $E_{\rm p,GRBM}=(137\pm8)$~keV.
The last case is that of 990718: clearly, the peak energy estimate by K06,
$E_{\rm p,BATSE}=(498\pm53)$~keV is much higher than ours,
$E_{\rm p,GRBM}=232^{+45}_{-34}$~keV. We are confident that our results are
more reliable. Indeed, while the GRBM spectrum covers the entire
time profile, so does not that of K06: they missed the first and the last
$\sim 40$~s of the overall profile. Missing the final soft tail of the light
curve biased the $E_{\rm p}$ estimate towards harder values.

\subsection{Time resolved spectra}
\label{sec:time_resolved}
We selected the GRBs whose time profiles have been sampled by multiple 128-s
time intervals with spectral coverage. In particular we focused on the most common
cases, i.e. when the total light curves split into two parts, called ``A'' and ``B''
(Fig.~\ref{f:example_000226}). We excluded
those events whose total fluence has been split more inhomogeneously than 20\%--80\%.
We ended up with a sample of 10 GRBs with reasonably well determined parameters
with the {\sc cpl} model. We added the case of 971110, which happened
to be covered by three intervals that collected comparable fluences and with
well determined parameters. Finally, we examined three very long GRBs that have
been sampled by several ($>3$) intervals and for which it was possible to extract at
least three meaningful spectra each (Sect.~\ref{sec:very_long}).
%
\begin{figure}
\centering
\includegraphics[width=8cm]{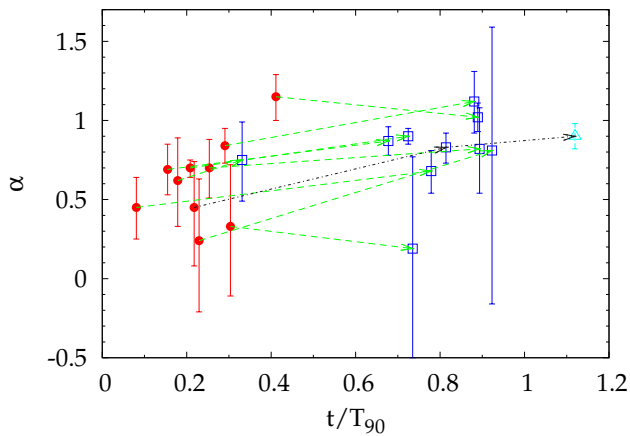}
\caption{Low--energy photon index as a function of time expressed
in units of $T_{90}$ for a subset of GRBs sampled by multiple spectral intervals.
Filled circles, empty squares and triangles correspond to the spectra
A, B and C, respectively. Each arrow tracks the evolution of a given GRB.}
\label{f:time_evol_alpha}
\end{figure}
%

\begin{figure}
\centering
\includegraphics[width=8.5cm]{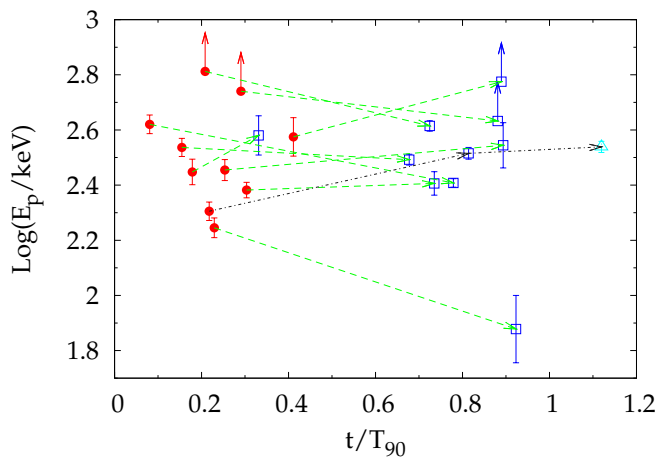}
\caption{Peak energy as a function of time expressed in units of $T_{90}$.
Same as Fig.~\ref{f:time_evol_alpha}.}
\label{f:time_evol_logEp}
\end{figure}

Figures~\ref{f:time_evol_alpha} and \ref{f:time_evol_logEp}
show the temporal evolution of the low--energy photon index and of the peak
energy, respectively, as a function of time. To account for the different durations
of the GRBs in the subsample, time is conveniently expressed in units of $T_{90}$ as
measured by F09. The time assigned to each interval was calculated as the
weighted-average over the 128 1--s bins, where the counts per bin in the light
curve were used as the weights: for each interval this procedure identifies the
time at which most of corresponding photons are observed. 
Filled circles and empty squares correspond to spectra ``A''
and ``B'', respectively. Dashed arrows connect points of the same GRB.
The case of 971110 is highlighted with dark arrows.
%
%
As can be seen, no global behaviour stands out.
While Figure~\ref{f:time_evol_alpha} suggests a marginal hard-to-soft evolution
of the photon index, the peak energy (Fig.~\ref{f:time_evol_logEp}) shows all
the possible cases compatibly with no standard evolution, in agreement with
early observations of GRBs from past experiments (e.g., \citealt{Kargatis94}; K06).
The variety of the peak energy evolution throughout the time profile of a GRB
is known to undergo a range of different behaviours: either tracking of the light
curve or a steady hard--to--soft evolution are observed (e.g. see \citealt{Peng09a}
and references therein). It must be pointed out that these results are derived from a
sample of bright GRBs and including fainter events could change the average
evolution of the spectral parameters.

\subsubsection{Very long GRBs}
\label{sec:very_long}
\begin{figure}
\centering
\includegraphics[width=8cm]{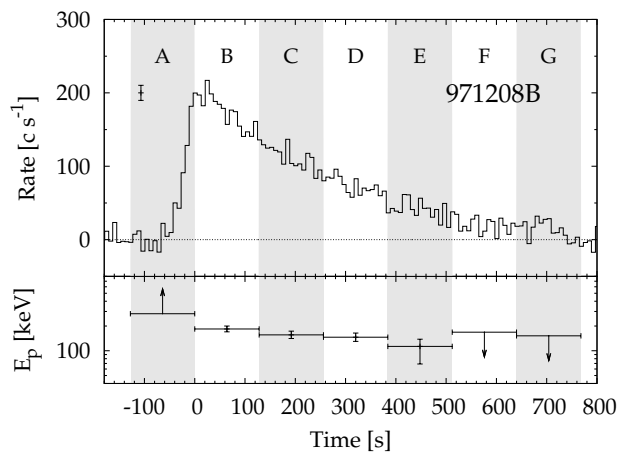}
\caption{{\em Top panel}: 40--700~keV time profile of 971208B.
{\em Bottom panel}: peak energy evolution. Shaded area identify
the different 128-s intervals over which an average spectrum was acquired.}
\label{f:971208B}
\end{figure}

\begin{figure}
\centering
\includegraphics[width=8cm]{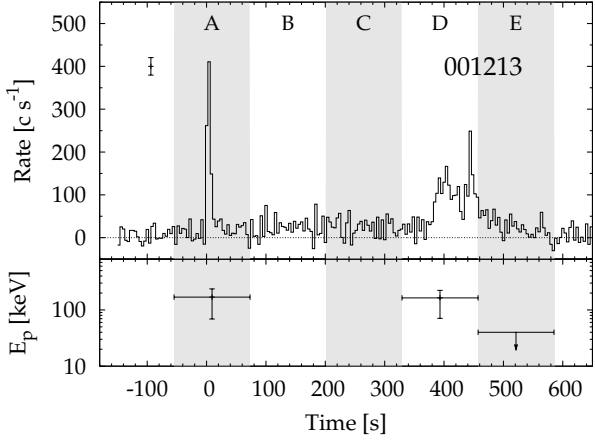}
\caption{GRB~001213. Same plot as Fig.~\ref{f:971208B}.}
\label{f:001213}
\end{figure}

\begin{figure}
\centering
\includegraphics[width=8cm]{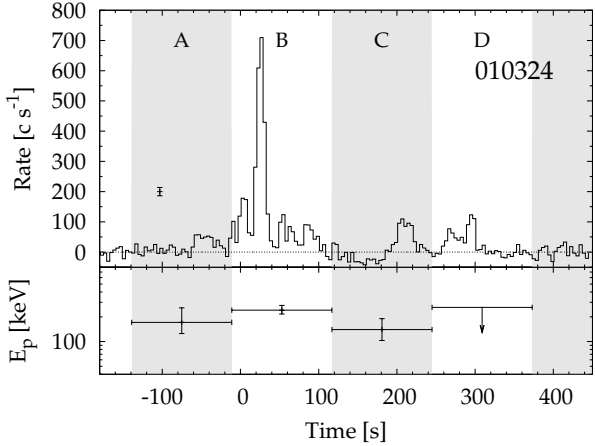}
\caption{GRB~010324. Same plot as Fig.~\ref{f:971208B}.}
\label{f:010324}
\end{figure}

We examined three of the longest GRBs of our set, that happened to be
sampled by several time intervals. For these GRBs we provide a more detailed
analysis of how $E_{\rm p}$ evolves with time compared with the time profile.
Figures~\ref{f:971208B}, \ref{f:001213} and \ref{f:010324} show 971208B,
001213 and 010324, respectively: each top panel shows the 40--700~keV time
profile with the typical error bar shown in the upper left corner,
while the bottom panel shows the peak energy of the corresponding
time intervals as a function of time.
%
%
These GRBs confirm the variety of $E_{\rm p}$ evolution compared with the
light curves: in the case of 971208B $E_{\rm p}$ steadily declined with time
even during the rise of the single, long-lasting pulse. By contrast, 
in the other cases it remains roughly constant throughout different emitting
episodes, followed by a final drop at the end of the prompt light curve.
%
%

%
%

\begin{figure}
\centering
\includegraphics[width=8cm]{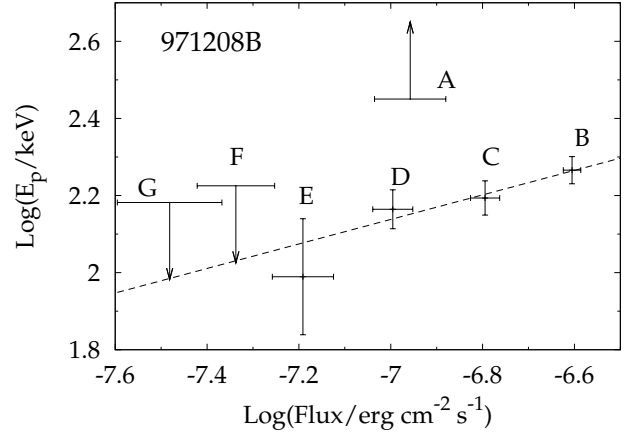}
\caption{Peak energy vs. average flux for 971208B. The dashed
line is the best-fitting power--law with a slope of $0.32\pm0.15$.
Labelled spectra are the same as Fig.~\ref{f:971208B}.}
\label{f:971208B_flux_ep}
\end{figure}

Thanks to its very long duration, the case of 9701208B offers the 
opportunity to study the relation between the average flux and $E_{\rm p}$
in each interval. The result is shown in Fig.~\ref{f:971208B_flux_ep}; the
dashed line shows the best-fitting power--law relation, as parametrised
by eq.~(\ref{eq:971208B_flux_ep}):
\begin{equation}
\log{\Big(\frac{E_{\rm p}}{{\rm keV}}\Big)}\ =\ m\
\log{\Big(\frac{{\rm Flux}}{{\rm erg\, cm^{-2}\, s^{-1}}}\Big)}\ +\ q
\label{eq:971208B_flux_ep}
\end{equation}
The best-fitting parameters computed over the four intervals with
measured $E_{\rm p}$, from ``B'' to ``E'', are $m=0.32\pm0.15$ and
$q=4.36\pm1.00$. Only spectrum ``A'' taken during the rise of the pulse
is not compatible with the power--law model and this was already
observed in other bursts with a broader energy coverage down to X--rays
for a sample of bursts detected with both the Wield Field Cameras (WFC)
and the GRBM aboard BeppoSAX (\citealt{Frontera10}; Frontera et al. in prep.).
This burst is also interesting because it belongs to the FRED (fast
rise exponential decay) class, a family of bursts with a single pulse
which are thought to be the building blocks of more complex time profiles
(e.g. \citealt{Norris96}). Similar results in the $E_{\rm p}$ evolution
of FRED GRBs are discussed by \citet{Peng09b}.
%
%

\section{Discussion and Conclusions}
\label{sec:disc}
We have analysed the spectral properties of the 185 brightest BeppoSAX/GRBM GRBs, 
using three different spectral models. The sample includes bright GRBs with a
threshold on fluence of $\Phi>4.4\times10^{-6}$~erg~cm$^{-2}$ in
the 40--700~keV band; as a consequence, no short duration GRB was selected. 
The GRBM data used consist of 240--energy channel spectra in the 40--700~keV range
continuously integrated over 128 s independently of the onboard trigger logic.
For this reason, the analysis mainly concerned the time--integrated
spectra of the GRBs; for a number of them, especially the very long ones,
it was possible to carry out the spectral analysis for a few contiguous
time intervals separately; these cases are referred to as time--resolved spectra.

About 35\% of the sample are best fit with a power--law model ({\sc pow}); 
the median value of the index is very close to 2.
The analogous fraction for the Fermi/GBM, the brightest BATSE and the
Swift known--$z$ GRBs samples is 30\%, 21\% and 38\%, respectively \citep{Nava10}.
The power--law index distribution is centred on $\alpha_{\rm pow}=1.86\pm0.32$
and agrees with other experiments \citep{Sakamoto08a,Vianello09}.
GRBs with $\alpha_{\rm pow}>2$ ($\alpha_{\rm pow}<2$) have a peak energy either
close to or below (above) the GRBM lower (upper) bound, so $E_{\rm p}\lesssim40$~keV
($E_{\rm p}\gtrsim700$~keV).

The typical long and bright GRBs are well fit with either a cut--off power--law
({\sc cpl}) or with a {\sc band} function; with the GRBM data
in none of the cases considered the latter model provided a significant
improvement with respect to the former. We also proved that within the
accuracy limits of these data, the two models provide consistent estimates for
both the low--energy photon index $\alpha$ and the peak energy $E_{\rm p}$,
although the {\sc cpl} model tends to overestimate $E_{\rm p}$ with respect
to the {\sc band} function.

For a sample of 28 GRBs commonly detected by both GRBM and BATSE and for which
K06 provided the results of the time-average spectral fitting, we carried
out a comparative analysis to establish possible discrepancies and to evaluate
the effects of measuring the same quantities with two different instruments.
The two sets of $\alpha$ and $E_{\rm p}$ substantially agree with one another,
except for a very few cases which we investigated and for which the main source
of discrepancy must be searched in the different time coverage.
A strong spectral evolution, observed for several GRBs, can explain why
different time intervals may yield significantly different results in the
spectral parameters. Specifically to $E_{\rm p}$, we modelled all these sources
of discrepancy in terms of an additional scatter of about 26\% between the GRBM
and BATSE $E_{\rm p}$ estimates. In practice, this has little impact on the
known correlations \citep{Amati02,Ghirlanda04} in which $E_{\rm p}$ is a key
observable.

The observed distribution of $E_{\rm p}$ peaks around 240~keV with a dispersion
of 0.2~dex, very similar to that of bright BATSE bursts (K06; \citealt{Nava10}).
Its narrowness is explained by the finite passband of the GRBM as well as of
the analogous experiments. That these selection effects connected with data
truncation affect the distribution, as discussed in Sect.~\ref{sec:res_cpl},
is directly proven by analogous studies carried out over broader energy ranges.
Indeed, in the X--ray domain \citep{Frontera00,Barraud03,Amati06,Pelangeon08,Sakamoto08b}
the number of X-ray rich GRBs and XRFs increases remarkably: as a result,
the $E_{\rm p}$ distribution forms a continuum over a correspondingly broad
energy band \citep{Sakamoto08b}. Moreover, the selected sample is not
representative of the entire population but only of the brightest end. 
A number of the GRBs included in our sample and best fit with a soft photon
index are likely to have $E_{\rm p}$ between the XRFs and the hardest GRBs.

The $\alpha$ distribution has its mode around 1 with a dispersion of $0.32$
(Fig.~\ref{f:alpha_hist}) very similar to the dispersion of $0.25$ found
for both low-- and high--energy indices by K06 on a sample of bright BATSE GRBs.
In this respect our results are in agreement with catalogues properties of
other experiments (\citealt{Preece00,Ghirlanda02}; K06; \citealt{Sakamoto08a,Nava10}).
For the GRBs with $E_{\rm p}<100\div150$~keV, $\alpha$ is poorly constrained
and is slightly biased towards soft values (Fig.~\ref{f:alpha_vs_logEp}).
As discussed in Sect.~\ref{sec:res_cpl}, this is due to an instrumental
effect which limits the capability of correctly measuring the low--energy
photon index when $E_{\rm p}$ lies close to the lower energy bound.
For such GRBs, the spectrum cannot reach the asymptotic slope at the
lower bound, so the value provided by the fitting procedure turns out to
be softer than the asymptotic one \citep{Lloyd00,Amati02}.
How close the observed slope can be with respect to its asymptote
depends on the spectrum itself and on its physical origin. 
For instance, assuming the validity of the synchrotron shock model, \citet{Lloyd00}
studied how the smoothness of the cutoff in the electron energy distribution
and the distribution of the pitch angle determine how quickly the asymptotic
value of $\alpha$ is reached within a given energy passband.

As also noted by K06, the $\alpha$ distribution does not exhibit any
clustering around characteristic values expected from various models:
2/3 for synchrotron with no cooling \citep{Katz94,Cohen97,Tavani96},
0 for jitter radiation expected in the case of synchrotron radiation in
highly non--uniform short--scale magnetic fields \citep{Medvedev00} 
and 3/2 for fast cooling synchrotron \citep{Ghisellini00}.

About 30\% of GRBs whose time--average spectrum is best fit with a {\sc cpl},
lie below the synchrotron death line of $\alpha=2/3$ (vertical dotted line
in Fig.~\ref{f:alpha_hist}). This fraction is comparable to that found in
BATSE GRB samples \citep{Preece98}.
On the other side of the distribution, the fast-cooling death line
represented by the limit $\alpha<3/2$ (vertical dashed line) is satisfied
by all GRBs. However, the synchrotron process has several problems:
the observed distribution around 1 is remarkably harder than 3/2 expected for a
population of cooling electrons in the fast regime. \citet{Ghisellini00} considered
several options to overcome this discrepancy (particle re--acceleration,
deviations from equipartition, quickly varying magnetic fields, adiabatic
losses) concluding that the prompt spectrum could not be the result of
ultrarelativistic electrons emitting synchrotron and inverse Compton radiation.
Alternatively, \citet{Lloyd00} found that spectra below the synchrotron
death line are still possibly produced via synchrotron, provided that
one assumes small pitch angles for the emitting electrons.
Other possible explanations include synchrotron self--absorption in the
X-ray \citep{Granot00}, the presence of a photospheric component and
pair formation \citep{Meszaros00,Ioka07}, synchrotron self--Compton upscattered
to X-rays from optical \citep{Panaitescu00}, time dependent acceleration
and radiation \citep{Lloyd02}, the decay of magnetic fields \citep{Peer06},
the Klein--Nishina effect on synchrotron self-Compton process \citep{Derishev01},
or continuous electrons acceleration as a consequence of plasma turbulence in
the post--shock region \citep{Asano09}.

We have confirmed the correlation between the observed peak energy $E_{\rm p}$ and
the 40--700~keV fluence $\Phi$ with a null hypothesis probability of $1.4\times10^{-7}$.
The slope of $0.21 \pm 0.06$ agrees with previous values obtained
on samples of BATSE GRBs: $0.28\pm0.04$ \citep{Lloyd00b}, $0.16\pm0.02$ \citep{Nava08}.
The extrinsic scatter is $\sigma_{\log{E_{\rm p}}}=0.13\pm0.02$ ~(Fig.~\ref{f:ep_vs_flu}). 
The observed slope is likely to be only barely affected by data truncation
and selection effects: in the literature, the same problem, affecting analogous samples
from other experiments like BATSE, was circumvented by means of non--parametric techniques
that had been set up to correct for data truncation. Furthermore, this was also done
through the analysis of subsamples of brighter GRBs, less affected by selection
effects due to the detector threshold, as is also the case of our sample.
Results on the $E_{\rm p}$--$\Phi$ correlation based on a proper treatment of these
effects provided similar results on the correlation slope (see \citealt{Lloyd00b}
and references therein).

That the slope is shallower than $\sim 0.5$, slope of the intrinsic
$E_{\rm p,i}$--$E_{\rm iso}$ relation \citep{Amati02}, is explained by a combination
of different factors: i) in the observer frame, observables are not redshift--corrected.
Due to the selection effects on $E_{\rm iso}$ with redshift (both observational
and evolutionary), the farthest GRBs have the largest $E_{\rm iso}$: this makes
the $E_{\rm p}$--$\Phi$ relation flatter than the intrinsic one. ii) The
difficulty of detecting GRBs with $E_{\rm p}$ values either close to or below
the lower bound of the GRBM passband of 40~keV, as suggested by \citet{Nava08}
in the case of BATSE, is an instrumental effect which limits the dynamical
range, thus leading to a flatter slope.
This seems to be confirmed by the results of \citet{Sakamoto08b}
who found a slope of $0.52\pm0.11$ for an extended sample of BATSE, HETE--II and
Swift GRBs, much more sensitive to lower $E_{\rm p}$ than the GRBM alone.
Although the extrinsic scatter found for the $E_{\rm p}$--$\Phi$ relation
is smaller than that of the intrinsic relation ($\sim0.2$),
this must be compared with the dynamical range along $E_{\rm p}$:
moving from the observer to the intrinsic plane, the ratio between scatter and
range along $E_{\rm p}$ significantly decreases, and the correlation becomes more
significant \citep{Amati09}.

In addition to the $\alpha$ distribution problem suffered by the synchrotron,
the dependence of $E_{\rm p}$ on the prompt emission radius $r$ is strong:
assuming a fixed magnetic field fraction $\epsilon_B$ of the central
luminosity $L$, it is $\epsilon_B\,L \simeq\Gamma^2\,B^2\,r^2\,c$ ($\Gamma$
is the bulk Lorentz factor of the baryonic outflow and $B$ the magnetic field).
If particles are accelerated at the shocks to random Lorentz factor $\gamma$,
then $E_{\rm p}$ is expected to be:
\begin{equation}
E_{\rm p}\ =\ \hbar\,\Gamma\,\gamma^2\ \frac{e\,B}{m_e\,c}\ =\
\frac{e\,\hbar}{c^{3/2}\,m_e\,}\ \frac{\gamma^2}{r}\ \sqrt{\epsilon_B\,L}
\simeq\ 100\ r_{14}^{-1}\, L_{52}^{1/2}\ \textrm{keV}
\label{eq:ep}
\end{equation}
where we adopted the notation $Q=Q_{n}\times10^{n}$ for a generic quantity $Q$
and assumed $\gamma\sim m_p/m_e$. Equation~(\ref{eq:ep}) naturally explains
the $E_{\rm p,i}$--$L$ relation, i.e. the time-resolved version of the
$E_{\rm p,i}$--$E_{\rm iso}$ relation \citep{Amati02}. However, 
relating $r$ to the minimum observed variability timescale, $r\sim c\,t_v\,\Gamma^2$,
implies the dependence of $E_{\rm p}$ on $\Gamma$, thus making the interpretation of
the $E_{\rm p,i}$--$L_{\rm iso}$ relation through eq.~(\ref{eq:ep}) troublesome.
The resultant dependence of $E_{\rm p}$ on $t_v$ is not observationally
established \citep{Lyutikov10}.
Recently, high-energy observations with Fermi of prompt-GeV correlated photons
of GRB~080916C \citep{Abdo09a} would imply $r\sim 10^{16}$~cm (based on the constraints
derived from the observations of GeV photons and minimum variability timescale in the
light curve prompt), while the observed $E_{\rm p}$ of $\sim 500$~keV \citep{Golenetskii08}
with time--resolved peaks up to a few MeV \citep{Abdo09a} is much larger than what expected
from eq.~(\ref{eq:ep}). The same arguments hold for other high--energy GRBs
detected with Fermi, such as GRB~090217A \citep{Ackermann10} and
GRB~090902B \citep{Abdo09b}.

Overall, the synchrotron shock model cannot account for the entire observed
phenomenology of the GRB prompt emission \citep{Kumar07,Kumar08,Kumar09}.
Alternatively to the fireball model, in which
most of the energy is initially bulk kinetic energy of a relativistic outflow
turning into radiation through shocks, electromagnetic models have also been
proposed in which the bulk energy is carried by magnetic fields and particle
acceleration occurs through magnetic dissipation instead of shocks
(\citealt{Lyutikov06} and references therein).

More generally, the magnetic energy content of the ejecta can be explored effectively
through early polarisation measurements, when most of the radiation comes
from the ejecta themselves rather than the shocked interstellar medium as observed
in the afterglow.
At optical wavelengths this is thought to be the case whenever a reverse shock, which
propagates through the ejecta, dominates the observed radiation during the $\gamma$--ray
prompt emission or immediately afterwards (e.g., \citealt{Zhang05}). 
This kind of measurements can discriminate between the possible different origins of
the magnetic field: either a large--scale magnetic field originating at the central
engine and carried forward by the ejecta, or a magnetic field generated in situ through the shocks.
Although this kind of measurements is still in its infancy and more data are required
to draw firm conclusions, early results on two recent bursts support the large--scale
magnetic field scenario within ejecta with comparable magnetic and kinetic energy contents
\citep{Mundell07,Steele09}.

Finally, for the longest GRBs it was possible to perform a time--resolved analysis
with temporal resolution bound to 128~s. We confirmed the absence of a general
evolution of the spectral parameters, especially $E_{\rm p}$, throughout the GRB
time profile: either a monotonic decline irrespective of the light curve or
no remarkable evolution at all. In the case of 971208B, an almost $10^3$--s long
FRED, we tracked the spectral evolution in the $E_{\rm p}$--flux observer plane,
finding it consistent with the $E_{\rm p,i}$--$L$ relation \citep{Yonetoku04,Ghirlanda05},
except for the spectrum of the pulse rise, clearly incompatible with it
(Fig.~\ref{f:971208B_flux_ep}). This behaviour appears to be naturally explained
in the context of synchrotron emission (eq.~\ref{eq:ep}), although the issues
mentioned above must be also considered.
Tracking the peak energy evolution over a broader energy band down to X--rays will
be crucial to better test the validity limits of this relation across individual
GRBs as well as samples of GRBs and, consequently, to gain insight on the nature
of the dominant emission process during the GRB itself. The importance of a broad
band coverage is already shown by the time--resolved combined spectral study
with WFC and GRBM aboard BeppoSAX on a sample of GRBs detected with both
instruments (\citealt{Frontera10}; Frontera et al. in prep.) as well as by the
broadband spectral analysis of X--ray flares detected with Swift \citep{Margutti10}.
These capabilities will be of key importance with future missions like
SVOM \citep{Dong10} and MIRAX \citep{Braga06,Braga10}.


\acknowledgements{}
We thank L. Nava for kindly supplying us with data of hers.
This work is supported by ASI contract ASI-INAF I/088/06/0
"Studio di Astrofisica delle Alte Energie".
We also thank the referee Vah\'e Petrosian for his useful comments, which
led to improvements in the paper.



\longtab{1}{
\begin{longtable}{rlrcrrrrl}
\caption{\label{tab:packets} Log of the 128-s energy spectra of the brightest GRBM bursts.
The SOD is measured in seconds of day of the on-ground trigger time (F09). The ID is a letter assigned to
each spectrum of a given GRB.}\\
\hline\hline
\#   & GRB         & SOD  & ID & Packet & $t_{\rm start}^{(a)}$ &  $t_{\rm stop}^{(a)}$ & $T_{90}$ & units\\
-   &  -         & (s)  & - & \# & (s) &  (s) & (s) & -\\
\noalign{\smallskip}
\hline
\noalign{\smallskip}
\endfirsthead
\caption{continued.}\\
\hline\hline\hline
\#   & GRB         & SOD  & ID & Packet & $t_{\rm start}^{(a)}$ &  $t_{\rm stop}^{(a)}$ & $T_{90}$ & units\\
-   &  -         & (s)  & - & \# & (s) &  (s) & (s) & -\\
\noalign{\smallskip}
\hline
\noalign{\smallskip}
\endhead
\hline
\endfoot
  1  &     960703  & 49374 & A &  29 &  -96.6 &   31.4 &   71 & 1 4 -\\
  1  &     960703  & 49374 & B &  30 &   31.4 &  159.4 &   71 & 1 4 -\\
  2  &    960707B  & 59186 & A &   2 &  -17.3 &  110.7 &   90 & 3 - -\\
  3  &    960723A  & 17163 & A &  10 &  -55.2 &   72.7 &   54 & 2 - -\\
  4  &    960805B  & 78957 & A &  14 & -120.3 &    7.7 &   14 & 1 3 -\\
  4  &    960805B  & 78957 & B &  15 &    7.7 &  135.7 &   14 & 1 3 -\\
  5  &     960806  & 80916 & A &   8 & -126.3 &    1.7 &   93 & 2 3 -\\
  5  &     960806  & 80916 & B &   9 &    1.7 &  129.7 &   93 & 2 3 -\\
\end{longtable}
  \begin{list}{}{} 
  \item[$^{\mathrm{(a)}}$] Time is referred to the on-ground trigger time SOD (third column).
  \item[$^{\mathrm{(b)}}$] Unavailable data packets.
  \item[$^{\mathrm{(c)}}$] Unknown start and end times.
  \item[$^{\mathrm{(d)}}$] Low signal.
  \item[$^{\mathrm{(e)}}$] Bad spectra.
  \end{list}   
}

\longtab{2}{
\begin{longtable}{rllccccccl}
\caption{\label{tab:results} Spectral fitting results of the time-integrated spectra for each bright GRB.
Spectral models used are: Band's model ({\sc Band}), the cut-off power--law ({\sc cpl}) and the simple
power--law ({\sc pow}). $\alpha$ and $\beta$ are the low- and high-energy photon indices ($\beta$ is not
defined for the {\sc cpl} and {\sc pow} models).
$E_{\rm p}$ is the peak energy of the $\nu\,F\,\nu$ spectrum. CAT is the catalogue ID which provided the
most accurate position for the GRB and is taken from \cite{Frontera09}. Frozen values are reported
among square brackets. For each GRB the best--fit model is marked with an asterisk.}\\
\hline\hline
\# GRB &  GRB &  model & $\alpha$ &  $\beta$ & $E_{\rm p}$ & $\Phi(40-700)$ & $\chi^2/{\rm dof}$ &  CAT &     
 comment\\
- &  - &  - & - &  - & (keV) & ($10^{-5}$ erg cm$^{-2}$) & - &  - &  -\\
\noalign{\smallskip}
\hline
\noalign{\smallskip}
\endfirsthead
\caption{continued.}\\
\hline\hline\hline
\# GRB &  GRB &  model & $\alpha$ &  $\beta$ & $E_{\rm p}$ & $\Phi(40-700)$ & $\chi^2/{\rm dof}$ &  CAT &     
 comment\\
- &  - &  - & - &  - & (keV) & ($10^{-5}$ erg cm$^{-2}$) & - &  - &  -\\
\noalign{\smallskip}
\hline
\noalign{\smallskip}
\endhead
\hline
\endfoot
  1  &    960703 & {\sc cpl}  & $1.45^{+0.33}_{-0.40}$ & - & $> 264$ & $ 1.23\pm 0.10$ & 0.77 &    4B 5526 & K06; only unit 1 used   \\
  1  &    960703 & {\sc pow} (*)  & $1.71^{+0.11}_{-0.11}$ & - & - & $ 1.26\pm 0.08$ & 0.78 &    4B 5526 & K06; only unit 1 used    \\
  2  &   960707B & {\sc cpl}  & $0.67^{+1.06}_{-1.68}$ & - & $> 180$ & $ 0.40\pm 0.10$ & 0.81 &    4B 5532 &       \\
  2  &   960707B & {\sc pow} (*)  & $1.67^{+0.27}_{-0.27}$ & - & - & $ 0.46\pm 0.07$ & 0.88 &    4B 5532 &       \\
  3  &   960723A & {\sc cpl} (*)  & $0.25^{+1.31}_{-1.93}$ & - & $120^{+34}_{-37}$ & $ 0.45\pm 0.09$ & 0.55 & 4B 5551   - &       \\
  3  &   960723A & {\sc pow}  & $2.17^{+0.26}_{-0.25}$ & - & - & $ 0.55\pm 0.09$ & 0.90 &    4B 5551 &       \\
  4  &   960805B & {\sc cpl} (*)  & $0.12^{+0.56}_{-0.69}$ & - & $  176^{ +13}_{ -12}$ & $ 2.68\pm 0.20$ & 1.06 &    - &       \\
  5  &    960806 & {\sc pow} (*)  & $1.80^{+0.18}_{-0.18}$ & - & - & $ 1.44\pm 0.14$ & 1.20 &    4B 5566 &       \\
\end{longtable}
  \begin{list}{}{} 
  \item[$^{\mathrm{(a)}}$] Unavailable data packets.
  \item[$^{\mathrm{(b)}}$] Bad spectra.
  \item[$^{(*)}$] Best fit model for a given GRB.
  \end{list}   
}

\longtab{3}{
\begin{longtable}{rlclccccccl}
\caption{\label{tab:results_time} Spectral fitting results of the time-resolved spectra for each bright GRB
that happened to be sampled by more 128-s time intervals. Spectral models are the same as in Table~\ref{tab:results}.
Intervals are the same as in Table~\ref{tab:packets}. Asterisks mark the best-fit model for the spectra that were fit
with different models.}\\
\hline\hline
\# GRB &  GRB &  Interval & model & $\alpha$ &  $\beta$ & $E_{\rm p}$ & $\Phi(40-700)$ & $\chi^2/{\rm dof}$ &  CAT &     
 comment\\
- &  - & - &  - & - &  - & (keV) & ($10^{-5}$ erg cm$^{-2}$) & - &  - &  -\\
\noalign{\smallskip}
\hline
\noalign{\smallskip}
\endfirsthead
\caption{continued.}\\
\hline\hline\hline
\# GRB &  GRB & Interval &  model & $\alpha$ &  $\beta$ & $E_{\rm p}$ & $\Phi(40-700)$ & $\chi^2/{\rm dof}$ &  CAT &     
 comment\\
- &  - & - &  - & - &  - & (keV) & ($10^{-5}$ erg cm$^{-2}$) & - &  - &  -\\
\noalign{\smallskip}
\hline
\noalign{\smallskip}
\endhead
\hline
\endfoot
  1  &    960703 & A & {\sc cpl} & $0.89^{+0.43}_{-0.48}$ & - & $  307^{+140}_{ -56}$ & $ 0.75\pm 0.06$ & 0.84 &  4B 5526 & \\
  1  &    960703 & B & {\sc pow} & $2.04^{+0.19}_{-0.19}$ & - &         -        & $ 0.47\pm 0.05$ & 1.29 &  4B 5526 & \\
  4  &   960805B & A & {\sc pow} (*) & $2.08^{+0.20}_{-0.20}$ & - &         -        & $ 0.92\pm 0.11$ & 1.03 & - & \\
  4  &   960805B & A & {\sc cpl} & $< 0.65$ & - & $  194^{ +31}_{ -25}$ & $ 0.71\pm 0.13$ & 0.68 & - & \\
  4  &   960805B & B & {\sc cpl} & $0.33^{+0.44}_{-0.48}$ & - & $  164^{ +13}_{ -14}$ & $ 1.96\pm 0.15$ & 0.85 & - & \\
  5  &    960806 & A & {\sc pow} & $1.96^{+3.03}_{-1.78}$ & - &         -        & $ 0.19\pm 0.10$ & 1.36 &  4B 5566 & \\
  5  &    960806 & B & {\sc cpl} & $0.58^{+0.68}_{-0.85}$ & - & $  231^{ +56}_{ -35}$ & $ 1.11\pm 0.13$ & 0.77 &  4B 5566 & \\
  6  &    960825 & A & {\sc cpl} & $0.33^{+0.39}_{-0.44}$ & - & $  241^{ +16}_{ -15}$ & $ 2.82\pm 0.15$ & 0.58 &  G & \\
\end{longtable}
  \begin{list}{}{} 
  \item[$^{\mathrm{(a)}}$] Low signal, no useful spectrum.
  \item[$^{\mathrm{(b)}}$] Unacceptable fit due to large $\chi^2$.
  \end{list}   
}


\begin{thebibliography}{}

\bibitem[Abdo et~al.(2009a)]{Abdo09a}
Abdo, A.A., Ackermann, M., Arimoto, M. et al. 2009a, Science, 323, 1688

\bibitem[Abdo et~al.(2009b)]{Abdo09b}
Abdo, A.A., Ackermann, M., Ajello, M. et al. 2009b, \apj, 706, L138

\bibitem[Ackermann et~al.(2010)]{Ackermann10}
Ackermann, M., Ajello, M., Baldini, M., et al. 2010, \apj, 717, L127

\bibitem[Amati et~al.(2002)]{Amati02}
Amati, L., Frontera, F., Tavani, M., et~al. 2002, \aap, 390, 81

\bibitem[Amati(2006)]{Amati06}
Amati, L. 2006, \mnras, 372, 233

\bibitem[Amati~et~al.(2009)]{Amati09}
Amati, L., Frontera, F., Guidorzi, C. 2009, \aap, 508, 173

\bibitem[Arnaud(1996)]{Arnaud96} 
Arnaud, K., A., 1996, in Astronomical Society of the Pacific Conference Series, 
Vol. 101, Astronomical Data Analysis Software and Systems V, 
ed. G. H. Jacoby \& J. Barnes, 17 

\bibitem[Asano \& Teresawa(2009)]{Asano09}
Asano, K. \& Teresawa, T. 2009, \apj, 705, 1714

\bibitem[Band et~al.(1993)]{Band93}
Band, D., Matteson, J., Ford, L., et~al. 1993, \apj, 413, 281

\bibitem[Barraud et al.(2003)]{Barraud03}
Barraud, C., Olive J.-F., Lestrade, J.P. et al. 2003, \aap, 400, 1021

\bibitem[Boella et~al.(1997)]{Boella97}
Boella, G., Butler, R. C., Perola, G. C., et~al. 1997, A\&AS,  122, 299

\bibitem[Braga~\&~Mej\'{\i}a(2006)]{Braga06}
Braga, J., Mej\'{\i}a, J. 2006, Proc.~SPIE, 6266, 62660M

\bibitem[Braga et~al.(2010)]{Braga10}
Braga, J., et~al. 2010, IJMPD, Proc. of 2nd Galileo-Xu Guangqi Meeting, held in Villa
Hanbury in July, 2010 , P. Chardonnet, L.-Z. Fang, R. Ruffini (Eds.), in press

\bibitem[Cabrera et~al.(2007)]{Cabrera07}
Cabrera, J. I., Firmani, C., Avila-Reese, V., et~al. 2007, \mnras, 382, 342

\bibitem[Calura et al.(2000)]{Calura00}
Calura, F., Rapisarda, M., Frontera, F., et~al., 2000,  Proc.~AIP, 526, 721

\bibitem[Campana et al.(2007)]{Campana07}
Campana, S., Guidorzi, C., Tagliaferri, et~al., 2007, 472, 395

\bibitem[Cohen et al.(1997)]{Cohen97}
Cohen, E., Katz, J.I., Piran, T. et~al. 1994, \apj, 488, 330

\bibitem[Derishev et~al.(2001)]{Derishev01}
Derishev, E.V., Kocharovsky, V.V., \& Kocharovsky Vl.V. 2001, \aap, 372, 1071

\bibitem[Dong et~al.(2010)]{Dong10}
Dong, Y.W., Wu, B.B., Li, Y.G., et al. 2010, Sc. Ch. G., 53, 40 (arXiv: 0907.2768)

\bibitem[Feroci et~al.(1997)]{Feroci97}
Feroci, M., Frontera, F., Costa, E. et~al. 1997, Proc.~SPIE, 3114, 186

\bibitem[Frontera et~al.(1997)]{Frontera97}
Frontera, F., Costa, E., Dal Fiume, D. et~al. 1997, A\&AS, 122, 357

\bibitem[Frontera et~al.(2000)]{Frontera00}
Frontera, F., Amati, L., Costa, E. et~al. 2000, \apjs, 127, 59

\bibitem[Frontera et al.(2009)]{Frontera09}
Frontera, F., Guidorzi, C., Montanari, E. et al. 2009, \apjs, 180, 192 (F09)

\bibitem[Frontera et al.(2010)]{Frontera10}
Frontera, F., Amati, L., Guidorzi, C., Landi, R., \& La Parola, V. 2010, \memsai,
81, 426

\bibitem[Ghirlanda et~al.(2002)]{Ghirlanda02}
Ghirlanda, G., Celotti, A., \&  Ghisellini, G. 2002, \aap, 393, 409 

\bibitem[Ghirlanda et~al.(2004)]{Ghirlanda04}
Ghirlanda, G., Ghisellini, G., \& Lazzati, D. 2004, \apj, 616, 331

\bibitem[Ghirlanda et~al.(2005)]{Ghirlanda05}
Ghirlanda, G., Ghisellini, G., Firmani, C., Celotti, A. \& Bosnjak, Z.,
2005, \mnras, 360, L45

\bibitem[Ghirlanda et~al.(2007)]{Ghirlanda07}
Ghirlanda, G., Nava, L., Ghisellini, G., Firmani, \aap, 466, 127

\bibitem[Ghisellini et~al.(2000)]{Ghisellini00}
Ghisellini, G., Celotti, A. \& Lazzati, D. 2000, \mnras, 313, L1

\bibitem[Golenetskii et~al.(2008)]{Golenetskii08}
Golenetskii, H., Aptekar, R., Mazets, E., et al. 2008, GCN Circ., 8258

\bibitem[Granot et~al.(2000)]{Granot00}
Granot, J., Piran, T. \& Sari, R. 2000, \apj, 534, L163


\bibitem[Guidorzi(2002)]{Guidorzi02}
Guidorzi, C. 2002, PhD thesis, Univ. of Ferrara

\bibitem[Guidorzi et~al.(2006)]{Guidorzi06}
Guidorzi, C., Frontera, F., Montanari, E. et al. 2006, \mnras, 371, 843

\bibitem[Heise et~al.(2001)]{Heise01}
Heise, J., in' t Zand, J., Kippen, R.M., \& Woods, P.M. 2001, in Gamma-Ray
Bursts in the Afterglow Era, ed. E. Costa, F. Frontera, \& J. Hjorth
(Berlin: Springer), 16

\bibitem[Ioka et~al.(2007)]{Ioka07}
Ioka, K., Murase, K., Toma, K. et al. 2007, \apj, 670, L77

\bibitem[Kargatis et~al.(1994)]{Kargatis94}
Kargatis, V.E., Liang, E.P., Hurley, K.C., et al. 1994, \apj, 422, 260

\bibitem[Kaneko et~al.(2006)]{Kaneko06}
Kaneko, Y., Preece, R.D., Briggs, M.S., Paciesas, W.S., Meegan, C.A.,
Band, D.L. 2006, \apjs, 166, 298 (K06)


\bibitem[Katz(1994)]{Katz94}
Katz, J.I. 1994, \apj, 432, L107

\bibitem[Kumar et~al.(2007)]{Kumar07}
Kumar, P., McMahon, E., Panaitescu, A., et al., 2007, \mnras, 376, L57

\bibitem[Kumar \& McMahon(2008)]{Kumar08}
Kumar, P. \& McMahon, E., 2008, \mnras 384, 33 

\bibitem[Kumar \& Narayan(2009)]{Kumar09}
Kumar, P. \& Narayan, R., 2009, \mnras, 395, 472

\bibitem[Lloyd \& Petrosian(1999)]{Lloyd99}
Lloyd, N.M. \& Petrosian, V. 1999, \apj, 511, 550

\bibitem[Lloyd \& Petrosian(2000)]{Lloyd00}
Lloyd, N.M. \& Petrosian, V. 2000, \apj, 543, 722

\bibitem[Lloyd, Petrosian \& Mallozzi(2000)]{Lloyd00b}
Lloyd, N.M., Petrosian, V., Mallozzi, R.S. 2000, \apj, 534, 227

\bibitem[Lloyd \& Petrosian (2002)]{Lloyd02}
Lloyd--Ronning, N.M. \& Petrosian, V. 2002, \apj, 565, 182

\bibitem[Lyutikov(2006)]{Lyutikov06}
Lyutikov, M. 2006, New J. Phys., 8, 119

\bibitem[Lyutikov(2010)]{Lyutikov10}
Lyutikov, M. 2010, Conf. Procs ``The Shocking Universe meeting'', Venice,
G.~Chincarini, P. D'Avanzo, R.~Margutti and R.~Salvaterra (Eds.),
SIF, Bologna, Vol. 102, 3 

\bibitem[Margutti et~al.(2010)]{Margutti10}
Margutti, R., Guidorzi, C., Chincarini, G., M.~G.~Bernardini, F.~Genet,
J.~Mao, F.~Pasotti, 2010, \mnras, 406, 2149

\bibitem[McBreen et~al.(2010)]{McBreen10}
Mc Breen, S., Kr\"uhler, T., Rau, A., et~al., 2010, \aap, 516, A71

\bibitem[Medvedev(2000)]{Medvedev00}
Medvedev, M.V. 2000, \apj, 540, 704

\bibitem[Meegan~et~al.(1992)]{Meegan92}
Meegan, C. A., Fishman, G. J., Wilson, R. B., et~al., 1992, \nat, 355, 143

\bibitem[Meszaros \& Rees(2000)]{Meszaros00}
Meszaros, P. \& Rees, M.J. 2000, \apj, 530, 292

\bibitem[Mundell~et~al.(2007)]{Mundell07}
Mundell, C. G., Steele, I. A., Smith, R. J., et~al. 2007, Science, 315, 1822 

\bibitem[Nava et al.(2008)]{Nava08}
Nava, L., Ghirlanda, G., Ghisellini, G., and Firmani, C. 2008, \mnras, 391, 629

\bibitem[Nava et al.(2010)]{Nava10}
Nava, L., Ghirlanda, G., Ghisellini, G., and Celotti, A. 2010, submitted
(arXiv:1004.1410)

\bibitem[Norris et al.(1996)]{Norris96}
Norris, J.P., Nemiroff, R.J., Bonnell, J.T. 1996, \apj, 459, 393

\bibitem[Paciesas et~al.(1999)]{Paciesas99}
Paciesas, W.S., Meegan, C.A., Pendleton, G.N., et~al., 1999, \apjs, 122, 465

\bibitem[Panaitescu \& Meszaros(2000)]{Panaitescu00}
Panaitescu, A. \& Meszaros, P. 2000, \apj, 544, L17

\bibitem[Papathanassiou(1999)]{Papathanassiou99}
Papathanassiou H., 1999, A\&AS, 138, 525

\bibitem[Pe'er \& Zhang(2006)]{Peer06}
Pe'er, A. \& Zhang, B. 2006, \apj, 653, 454

\bibitem[Pelangeon et~al.(2008)]{Pelangeon08}
Pelangeon, A., Atteia, J-L., Nakagawa, Y.E., et al. 2008, \aap, 491, 157 

\bibitem[Peng et~al.(2009a)]{Peng09a}
Peng, Z.Y., Ma, L., Lu, R.J., et al. 2009a, \na, 14, 311

\bibitem[Peng et~al.(2009b)]{Peng09b}
Peng, Z.Y., Ma, L., Zhao, X.H. et al. 2009b, \apj, 698, 417

\bibitem[Preece et~al.(1998)]{Preece98}
Preece, R.D., Briggs, M.S., Mallozzi, R.S., Pendleton, G.N., Paciesas, W.S., Band, D.L. 
1998, \apj, 506, L23

\bibitem[Preece et~al.(2000)]{Preece00}
Preece, R.D., Briggs, M.S., Mallozzi, R.S., Pendleton, G.N., Paciesas, W.S.
2000, \apjs, 126, 19

\bibitem[Sakamoto et~al.(2005)]{Sakamoto05}
Sakamoto, T., Lamb, D.~Q., Kawai, N., et~al. 2005, \apj, 629, 311

\bibitem[Sakamoto et~al.(2008a)]{Sakamoto08a}
Sakamoto, T., Barthelmy, S.D., Barbier, L. et al. 2008a, \apjs, 175, 179

\bibitem[Sakamoto et~al.(2008b)]{Sakamoto08b}
Sakamoto, T., Hullinger, D., Sato, G. et al. 2008b, \apj, 679, 570

\bibitem[Sari et~al.(1998)]{Sari98}
Sari, R., Piran, T., \& Narayan, R. 1998, \apj, 497, L17

\bibitem[Steele~et~al.(2009)]{Steele09}
Steele, I. A., Mundell, C. G., Smith R. J., Kobayashi, S.,
Guidorzi, C. 2009, \nat, 462, 767

\bibitem[Tavani (1996)]{Tavani96}
Tavani, M., 1996, \apj, 466, 768 

\bibitem[Vianello et~al.(2009)]{Vianello09}
Vianello, G., G\"otz, D., Mereghetti, S. 2009, \aap, 495, 1005

\bibitem[Yonetoku et~al.(2004)]{Yonetoku04}
Yonetoku, D., Murakami, T., Nakamura, T., et al. 2004, \apj, 609, 935

\bibitem[Zhang~\&~Kobayashi(2005)]{Zhang05}
Zhang, B., Kobayashi, S. 2005, \apj, 628, 315

\end{thebibliography}
\end{document}